\documentclass[useAMS,usenatbib]{mn2e}
\usepackage{graphicx}
\usepackage{epstopdf}
\usepackage[figuresright]{rotating}
\usepackage{subfigure,color}
\usepackage{longtable}
\usepackage{xspace}

\newcommand{\ion}[2]{{\textrm{#1}}\,{\textrm{\sc #2}}}

\title[Ne abundance determinations in star-forming regions]{Optical and
mid-infrared neon abundance determinations in star-forming regions}
\author[Dors et al.]
{Oli L. Dors Jr.$^{1}$\thanks{E-mail:olidors@univap.br}, Guillermo
  F. H\"agele$^{2,3}$, M\'onica V.\ Cardaci$^{2,3}$, 
  \newauthor Enrique P\'erez-Montero$^{4}$, \^Angela C. Krabbe$^1$, Jos\'e M. V\'ilchez$^{4}$, 
 Dinalva A. Sales$^5$, 
  \newauthor Rog\'erio Riffel$^5$, Rogemar A. Riffel$^6$\\
$^1$ Universidade do Vale do Para\'iba, Av. Shishima Hifumi, 2911, Cep
12244-000, S\~ao Jos\'e dos Campos, SP, Brazil\\
 $^2$ Consejo Nacional de Investigaciones Cient\'ificas y T\'ecnicas (CONICET), Argentina.\\
$^3$Facultad de Ciencias Astron\'omicas y Geof\'{\i}sicas, Universidad Nacional de La Plata, Paseo del Bosque s/n, 1900 La Plata
, Argentina.\\
$^4$ Instituto de Astrof\'{\i}sica de Andaluc\'{\i}a (CSIC), PO Box 3004, 18080 Granada, Spain. \\ 
$^5$ Universidade Federal do Rio Grande do Sul, IF, CP 15051, Porto Alegre 91501-970, RS, Brazil.\\
$^6$ Universidade Federal de Santa Maria, Av. Roraima 1000, Cep 97105-900, Santa Maria, Brazil\\}

\begin{document}

\date{Accepted- 2011 April 28. Received -2011 February 18.}

\pagerange{\pageref{firstpage}--\pageref{lastpage}} \pubyear{2011}

\maketitle

\label{firstpage}

\begin{abstract}
We employed observational spectroscopic data of star-forming regions
compiled from the literature and photoionization models to analyse the
neon ionic abundances obtained using both optical and mid-infrared
emission-lines. Comparing
$\rm Ne^{++}/H^{+}$ ionic abundances from distinct methods, we found that, in
average, the abundances obtained via IR emission-lines are higher than those 
obtained via optical lines by a factor of 4. 
  Photoionization models with abundance variations along the radius of
the hypothetical nebula  provide a possible explanation for
a large part of the  difference between ionic abundances via optical and
infrared emission-lines. 
Ionization Correction Factor (ICF) for the neon is obtained from direct
determinations of ionic fractions using infrared emission-lines.  
 A constant  Ne/O  ratio ($\rm \log Ne/O \approx -0.70$)  for a large range of
metallicity, independently of the ICF used to compute the neon
total abundance is derived.  
 \end{abstract}

\begin{keywords}
galaxies: general -- galaxies: evolution -- galaxies: abundances --
galaxies: formation-- galaxies: ISM
\end{keywords}


\section{Introduction}


Gas phase metallicity determinations in star-forming regions (SFRs) 
have been used
for the knowledge of the cosmic evolution  of galaxies as well as to provide 
observational constraints on model parameters of galaxy formation.
 
In particular, the oxygen is the element most widely used for this purpose,
because prominent emission lines from their main ionic stages are present in
the optical spectra of SFRs
(e.g. \citealt{hagele08}; \citealt{kennicutt03}, \citealt{krabbe06}). 
For the rest of elements with bright emission-lines (e.g. N, S, Ne, Ar)
not all their ionic stages are observed in the optical spectrum, so their
total abundances can only be calculated by means of Ionization Correction
Factors (ICFs), proposed by \citet{peimbert69}. 

Among the $\alpha$-elements, Ne results of great importance 
as it is one of the noble gases and it does not
combine with itself or with other chemical species in the formation of
molecules and, thus, dust grains. This makes the abundance determination 
of this element quite suitable for the study
of the chemical evolution of SFRs as it does not
depend on depletion factors onto dust grains.
It is accepted that Ne and O are mainly produced in
stars more massive than 10 $M_{\odot}$ (see \citealt{woosley95}). Therefore, it is
expected that abundances of Ne and O should closely trace one to the other 
\citep{crockett06}, and a constant Ne/O abundance ratio over a wide range
of O/H abundance must be found. However, recent deep spectroscopy of a large
sample of low-metallicity emission-line galaxies by \citet{guseva11} revealed
a slight increase of Ne/O with O/H. Similar results were also found by
\citet{wang08} using O and Ne abundances of a sample of planetary nebulae and 
 \ion{H}{ii} regions. In opposite, \citet{perez07} derived a new ICF 
for Ne using a photoionization model grid 
finding for the high metallicity regime a slight decrease in Ne/O with O/H for a large sample of SFRs.
 
In any case, the derivation of the total Ne abundance has showed to be uncertain due to the difficulties estimating its ICF. 
The estimation of the total Ne abundance in the optical is performed via the observation
of the [\ion{Ne}{iii}]$\lambda$3869 \AA\ emission-line, which is 
mainly present in SFRs with a 
high-ionization degree,  and usually assuming an ICF($\rm Ne^{++}$) that 
is a function of the O/$\rm O^{++}$ ratio derived also from optical emission-lines.
However, the optical [\ion{Ne}{iii}] line is rarely observed in metal-rich objects (see \citealt{bresolin05}), 
where most of the Ne is in the form of $\rm Ne^{+}$,
so the ionization corrections are not well constrained in this regime \citep{kennicutt03}.
This problem can be alleviated by the use of infrared fine-structure lines \citep{vermeij02a}.
On this context, both the {\em Infrared Space Observatory} ({\em ISO}, \citealt{kessler96}) 
and the {\em Spitzer Space Telescope} have
allowed the observation of
bright mid-infrared emission-lines in SFRs, such as [\ion{Ne}{ii}]12.81$\mu$m and [\ion{Ne}{iii}]15.56$\mu$m,
and the determination of the total Ne abundance simply
summing the corresponding ionic fractions. For instance, \citet{vermeij02a} used optical and mid-IR data 
of \ion{H}{ii} regions located in the Magellanic Clouds 
\citep[see also][]{vermeij02} to determine O, S, and Ne abundances in these
objects. These authors found large discrepancies between $\rm Ne^{++}$
abundances obtained via infrared lines with those via optical temperature
determinations and pointed out that the classical approximation, ICF($\rm
Ne^{++}$)=O/$\rm O^{++}$ \citep{peimbert69}, seems to underestimate the true
Ne abundance (see also \citealt{perez07}). However, due to large scatter and
uncertainty in the $\rm Ne^{++}$ ionic fraction of their data, no conclusion
could be drawn about the reliability of the optical ICF for Ne. The same
results were obtained by \citet{kennicutt03}, who combined the ionic
abundances derived by \citet{vermeij02a} and by \citet{willner02} for
\ion{H}{ii} regions located in M\,33. However, they were not able to derive a
functional form for the ICF($\rm Ne^{++}$), due to the large scatter of the
available data and the very few measurements for low-ionization regions.
Fortunately, a large number of mid-IR and optical data of SFRs are currently
available in the literature, which enables a  direct determination of the 
optical ICF for Ne and a comparison among abundances obtained from distinct
methods, yielding a more reliable conclusion about the total Ne abundance.

The main goal of this paper is to examine the Ne abundance of SFRs, investigating
the abundance discrepancy found
by the use of different methods and obtaining a reliable ICF in the 
optical for this element.
To do that, we compiled observed optical and infrared
emission-line fluxes from the literature to derive the Ne ionic abundances. A photoionization model grid
 was also used in this analysis. The paper is organized as follows.
  In Section~\ref{obs} we describe the observational data used along the paper and
 an analysis of the selected sample is performed. A description
 of the photoionization models  is given in Sect.~\ref{mod}.
 The procedures to determine the  ionic abundances 
are presented in Sect.~\ref{ion}. 
A comparison between neon ionic
abundances obtained from distinct methods is given in Sect.~\ref{compneon}.
In Sect.~\ref{sicf}  the ICFs  computed for $\rm Ne^{++}$ and applied to the
data set are presented. The conclusions of the outcomes are
given in Sect.~\ref{conc}.

\section{Observational data}
\label{obs}

\subsection{Sample description}

\begin{figure}
\centering
\includegraphics[angle=-90,width=0.95\columnwidth]{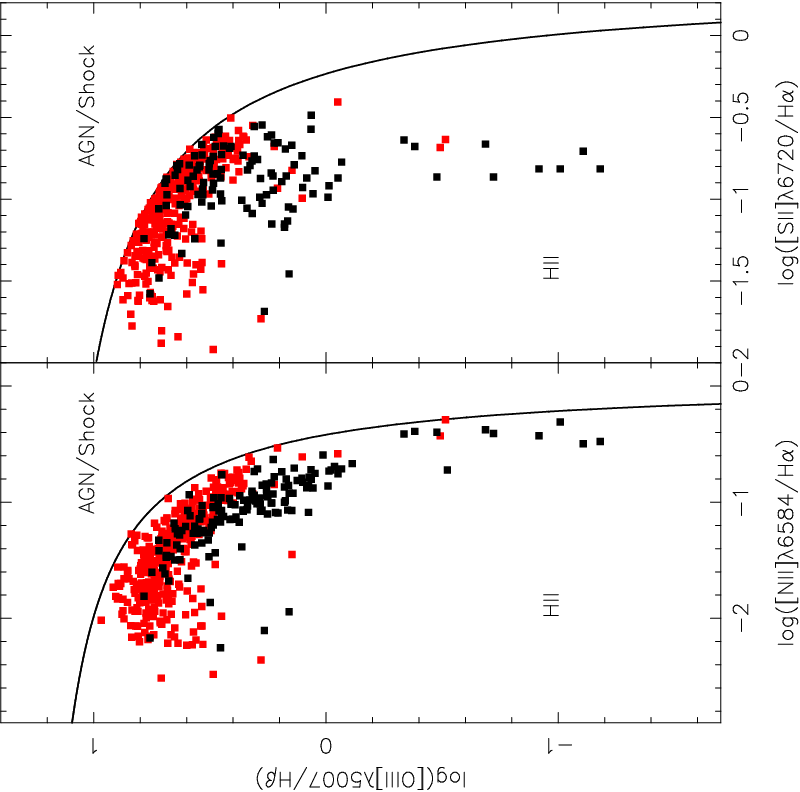}
\caption{ Diagnostic diagrams: [\ion{O}{iii}]$\lambda$5007/H$\beta$ vs.\
[\ion{N}{ii}]$\lambda$6584/H$\alpha$ (left) and
[\ion{O}{iii}]$\lambda$5007/H$\beta$ vs.\ 
[\ion{S}{ii}]($\lambda$6716+$\lambda$6731)/H$\alpha$ (right).
Solid lines, taken from \citet{kewley06}, separate objects ionized by massive
stars from the ones containing active nuclei and/or shock excited
gas as indicated.  Black squares represent \ion{H}{ii}Rs 
and red squares  the ELGs compiled from the literature.}
\label{fdisc1}
\end{figure}

Optical and infrared emission-line fluxes of a sample of emission-line
galaxies (ELGs) -- including \ion{H}{ii} galaxies
(\ion{H}{ii}Gs), Wolf Rayet galaxies
(WRGs), Blue Compact Galaxies (BCGs) and Dwarf Galaxies (DGs) -- and \ion{H}{ii}
regions (\ion{H}{ii}Rs) were compiled
from the literature. Their emission-line intensities were already reddening
corrected   in the works from which we take the data.   For most of
the data this was done by using the ratio $\rm H\alpha/H\beta$ and in some
cases, where $\rm H\alpha$  was not present in the spectra, the reddening
coefficient was obtained from another hydrogen lines (e.g. data of
\citealt{crockett06}). 

The selection criterion was the presence of flux measurements 
of the optical [\ion{O}{ii}]$\lambda$3727, [\ion{Ne}{iii}]$\lambda$3869, H$\beta$,
and [\ion{O}{iii}]$\lambda$5007 emission-lines.
 We also compiled the line intensities of [\ion{O}{iii}]$\lambda$4363,
H$\alpha$, [\ion{N}{ii}]($\lambda$5755,$\lambda$6548,$\lambda$6584) 
and [\ion{S}{iii}]($\lambda$6717,$\lambda$6731) when they were available.
Regarding mid-infrared emission-lines, we selected 50 ELGs
and 93 \ion{H}{ii}Rs located in the Magellanic Clouds,
M\,101, M\,33 and M\,83 whose [\ion{Ne}{ii}]12.81$\mu$m and [\ion{Ne}{iii}]15.56$\mu$m 
emission-lines were measured.
 The optical data were obtained by the use of long-slit, echelle and
 multi-object spectroscopy (MOS),
while the infrared data were obtained with the {\it ISO} and 
{\it Spitzer Space Telescope}.

In the optical, we used the [\ion{O}{iii}]$\lambda$5007/H$\beta$ vs.
[\ion{N}{ii}]$\lambda$6584/H$\alpha$ and   [\ion{O}{iii}]$\lambda$5007/H$\beta$ vs.
[\ion{S}{ii}]($\lambda$6716+$\lambda$6731)/H$\alpha$ diagnostic diagrams (Figure~\ref{fdisc1})  to distinguish
objects ionized  by massive stars from those containing 
an active galactic nucleus (AGN) and/or gas excited by shocks. To separate the
distinct classes of objects, we used  the criteria
proposed by \citet{kewley06}, where all objects with log[\ion{O}{iii}]$\lambda$5007/H$\beta < 0.61$/[log[\ion{N}{ii}]$\lambda$6584/H$\alpha$ -0.05] +1.3
and log[\ion{O}{iii}]$\lambda$5007/H$\beta < 0.72$/[log[\ion{S}{ii}]$\lambda$(6717+$\lambda$6731)/H$\alpha$ -0.32] +1.3
have massive stars as their main ionization mechanism. From this analysis we
discarded 15 objects yielding a sample with 522 ELGs and  212 \ion{H}{ii}Rs.

Due to the non-homogeneity of the compiled data, we divided the sample into
five groups according to the available emission-lines measured for each
object: 
 \begin{itemize}
 \item Group A - optical emission-lines, including the ones needed to directly estimate the electron
  temperatures and abundances. 
 \item Group B -  [\ion{O}{ii}]$\lambda$3727,
  [\ion{Ne}{iii}]$\lambda$3869, H$\beta$, [\ion{O}{iii}]$\lambda$5007 (the
  optical selection criterion) and [\ion{O}{iii}]$\lambda$4363.
 \item Group C - optical emission-lines except the
  ones sensitive to the temperature: [\ion{O}{iii}]$\lambda$4363, 
  [\ion{O}{ii}]$\lambda$7325 and [\ion{N}{ii}]$\lambda$5755. 
 \item Group D - only the [\ion{O}{ii}]$\lambda$3727,
  [\ion{Ne}{iii}]$\lambda$3869, H$\beta$, and [\ion{O}{iii}]$\lambda$5007
  emission-lines (just fulfilling the optical selection criterion). 
 \item Group E - the IR emission-lines.
 \end{itemize}
Table~\ref{tab1} lists the bibliographic references of the sample, the number
of objects taken from each work, their nature, the observational technique
used during the data acquisition, and the group(s) to which the data belong.
  The objects in the A, B and E groups allow the computation of
the ionic abundances and the ICFs from several methods used along the paper,
while the ones in the C and D groups only enable a more accurate comparison
between the data and the photoionization models to obtain a theoretical neon ICF.

 \begin{figure}
\centering
\includegraphics[angle=-90,width=8cm]{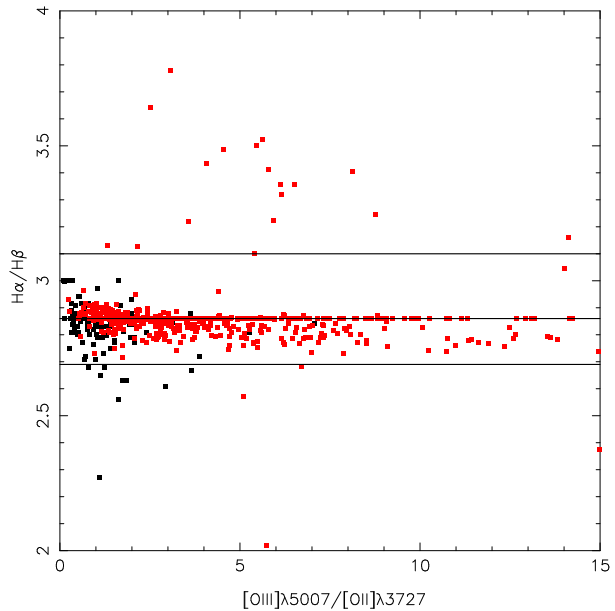}
\caption{H$\alpha$/H$\beta$ vs.
 [\ion{O}{iii}]$\lambda$5007/[\ion{O}{iii}]$\lambda$3727.  
 Black squares represent \ion{H}{ii}Rs 
and red squares  the ELGs compiled from the literature.   All data were 
 reddening corrected   in the works from which we taken the data.
The lines represent the theoretical values for the low
 density recombination case B Balmer decrement 
 H$\alpha$/H$\beta$=\,3.10, 2.86 and 2.69 for temperatures of 5\,000, 10\,000
 and 20\,000 K, respectively (\citealt{osterbrock89, storey95}).}
\label{f9}
\end{figure}

To investigate the discrepancy between the neon ionic abundances from optical
and infrared emission-lines, we selected the objects in our sample for which we
can get a direct estimation of, at least, an electron temperature. 
They are those belonging to the optical groups A and B, and also to the IR 
group E. The sub-sample contains 23 objects and they are listed in
Table~\ref{tab1a} together with the corresponding optical and IR
references. It is worth to emphasize that only for the objects in this 
sub-sample we can estimate the oxygen and neon ionic abundances via optical
lines and the neon ones via infrared lines. 

For the \ion{H}{ii}Rs located in the Magellanic Clouds, we converted the
Br$\beta$ fluxes, directly measured from the observations by \citet{vermeij02a},
in H$\beta$ fluxes using the emissivities of \citet{storey95}. We did
not use the Hu$\alpha$ emission line at 10.52\,$\mu$m, even though it is the
closest hydrogen line to the mid-IR Ne ones, because for our compiled sample it
is generally blended with an H$_2$ emission line. For the
remaining sources we considered the observed H$\beta$ fluxes. A similar procedure
was done by \citet{wu08} in their study of elemental abundances of BCGs
using $Spitzer$ observations.

\begin{table*}
\centering
\caption{Bibliographic references for the compiled sample.}
\label{tab1}
\vspace{0.2cm}
\begin{tabular}{@{}llcc ccc@{}}
\hline
ID &Reference              &       Object type &     Number  &  Technique   & Group \\	 
\hline
1 &\citet{lee04}              & KISS galaxies &	 13 & long-slit          & A \\
2 &\citet{vilchez03}          &	DG            &   4 & long-slit          & A \\
3 &\citet{hagele06}           & \ion{H}{ii}G  &   3 & long-slit          & A \\
4 &\citet{hagele08}           & \ion{H}{ii}G  &   6 & long-slit          & A \\
5 &\citet{hagele11}           & \ion{H}{ii}G  &   2 & long-slit          & A \\
6 &\citet{skillman89}         & DG            &   2 & long-slit          & A \\
7 &\citet{thuan05}            & BCG           &  26 & long-slit          & B \\
8 &\citet{masegosa94}         & \ion{H}{ii}G  &  99 & long-slit          & A \\
9 &\citet{papaderos08}        & \ion{H}{ii}G  &  24 & long-slit          & A \\
10&\citet{kobulnicky99}       & \ion{H}{ii}G  &   7 & long-slit          & A \\
11&\citet{izotov01}           & BCG           &   2 & long-slit          & A \\
12&\citet{izotov04}$^{\rm a}$  & BCG           &  33 & long-slit          & A \\
13&\citet{izotov98}$^{\rm a}$  & BCG           &  17 & long-slit          & A \\
14&\citet{izotov06c}$^{\rm a}$ & \ion{H}{ii}G  &   1 & long-slit          & A \\
15&\citet{izotov97}$^{\rm a}$  & BCG           &  28 & long-slit          & A \\
16&\citet{izotov94}$^{\rm a}$           & BCG           &  10 & long-slit          & A \\
17&\citet{izotov06b}          & BCG           & 125 & long-slit          & A \\
18&\citet{bergvall02}$^{\rm a}$& BCG           &   3 & long-slit          & A \\
19&\citet{guseva00}$^{\rm a}$  & WRG           &   8 & long-slit          & A \\
20&\citet{guseva11}           & \ion{H}{ii}G  &  59 & long-slit          & A \\
21&\citet{guseva07}$^{\rm a}$  & \ion{H}{ii}G  &  33 & long-slit          & B \\
22&\citet{lopez09}            & WRG           &  17 & long-slit          & A \\
23&\citet{kobulnicky97}       & \ion{H}{ii}R  &   3 & long-slit          & A \\
24&\citet{garnett97}          & \ion{H}{ii}R  &   7 & long-slit          & A \\
25&\citet{kennicutt03}        & \ion{H}{ii}R  &  20 & long-slit/echelle & A \\
26&\citet{vanzee98}           & \ion{H}{ii}R  &  58 & long-slit          & C \\
27&\citet{vilchez88}          & \ion{H}{ii}R  &   4 & long-slit          & A, C \\
28&\citet{kwitter81}          & \ion{H}{ii}R  &   6 &      slit          & A, C \\
29&\citet{bresolin09}         & \ion{H}{ii}R  &  28 & MOS                & A, C \\
30&\citet{lees04}             & \ion{H}{ii}R  &  15 & long-slit          & A, C \\
31&\citet{bresolin04}         & \ion{H}{ii}R  &   8 & MOS                & C \\
32&\citet{mccall85}           & \ion{H}{ii}R  &  20 & long-slit          & A, C, D \\
33&\citet{vanzee00}           & \ion{H}{ii}R  &   2 & long-slit          & A \\
34&\citet{bresolin11}         & \ion{H}{ii}R  &  19 & MOS                & D \\
35&\citet{esteban09}          & \ion{H}{ii}R  &   9 & long-slit          & A, C \\
36&\citet{crockett06}         & \ion{H}{ii}R  &  10 & long-slit          & B,  D \\
37&\citet{torrespeimbert89}$^{\rm a}$ & \ion{H}{ii}R &   3 & long-slit    &  A, C \\
38& \citet{engelbracht08}     &	\ion{H}{ii}G  &  41 & $Spitzer$          &  E \\
39&\citet{wu08}               &	      BCG     &	  9 & $Spitzer$          &  E \\
40&\citet{gordon08}           &	\ion{H}{ii}R  &	  7 & $Spitzer$          & E \\
41& \citet{vermeij02}$^{\rm a}$         &	\ion{H}{ii}R  &	  7 & {\em ISO}          & A, C, E \\
42& \citet{lebouteiller8}     & \ion{H}{ii}R  &  31 & $Spitzer$          & E \\
43& \citet{rubin08}           & \ion{H}{ii}R  &  24 & $Spitzer$          & E \\
44& \citet{rubin07}           & \ion{H}{ii}R  &  24 & $Spitzer$          & E \\
 \hline\noalign{\smallskip}
\end{tabular}
\begin{minipage}[c]{2\columnwidth}
 $^{\rm a}$ Papers from which the H$\beta$ fluxes were
 obtained to compute neon ionic abundances. 
\end{minipage}
\end{table*}

 \subsection{Analysis of the sample}
 \label{bias}
 
Along the years several authors have compiled spectroscopy data of
star-forming regions from the literature in order to derive correlations
between macroscopic properties of these objects (e.g. \citealt{pilyugin04}),
or have suggested oxygen abundance calibrations using strong emission-lines
(e.g.\ \citealt{perez09, nagao06}). However, the use of data sets
obtained with different instrumentation and observational techniques, and
including different objects such as whole galaxies, galaxy nuclei and
individual \ion{H}{ii}Rs, 
i.e.\ non-homogeneous data, could lead to biased results. 
In what follows we investigate the possible sources of biases introduced in
the present study due to the heterogeneity of our data set. 

\subsubsection{Aperture effects}

Differences between the   measurement  apertures used in the   optical and IR
observations could contribute significantly to the discrepancies found
between optical and IR abundance determinations \citep{vermeij02a}
 because  many physical properties (e.g., stellar populations, metallicity, extinction)
of galaxies vary with galactocentric radius  \citep{moustakas06a}
or along  nebulae \citep{oey00}. \citet{kewley05} presented a detailed analysis
of the effect of aperture size  on the star formation rate, metallicity, and reddening determinations for galaxies
selected from the Nearby Field Galaxy Survey. They found that  systematic and random errors from aperture effects
can arise if fibers   capture less than 20\% of the galaxy light.
Most of the SFRs in our sample can be treated as point sources   and almost
  all the object extensions are observed, therefore,
for these objects this effect is negligible. For the spatially resolved Magellanic \ion{H}{ii}Rs
in our sample (see Table \ref{tab1a}) this effect could play a more important
role as was already pointed out by \citet{vermeij02a}. 
However, since our aim is to study the discrepancy found for the Ne$^{++}$
ionic abundances and this ion is not expected to be formed in the extended low-surface
brightness zones around SFRs, aperture effects neither represent an important factor.

\subsubsection{Reddening correction}
\label{red}

There are several ways to determine the reddening correction to be applied to
the observed emission-line fluxes depending on the available information and
the ad hoc assumptions. For example, the use of different ratios involving
hydrogen emission-line fluxes, different extinction curves, and the use of a
wide spectral range. All these facts could introduce some biases in our
results. 
To investigate this effect, in Fig.~\ref{f9} we plotted the H$\alpha$/H$\beta$
ratios against the ionization degree parametrized by the 
[\ion{O}{iii}]$\lambda$5007/[\ion{O}{iii}]$\lambda$3727 ratio
\cite[e.g.][]{dors03,morisset04} for the objects belonging to our sample. 
In this figure we can see very few points ($\sim 3$\% ELGs and  $\sim 6$\% \ion{H}{ii}Rs) out of the range of the
expected values of H$\alpha$/H$\beta$ ratio considering   the theoretical values of \citet{storey95}  for temperatures between
5000 and 20000\,K,   typical values of star-forming regions. 
Thus, we can conclude that differences
in the reddening correction does not affect any statistical result obtained in
this work. 

 \begin{table}
\centering
\caption{Objects with neon abundances determined via optical and
infrared emission-line intensities.}
\label{tab1a}
\vspace{0.2cm}
\begin{tabular}{@{}lcc@{}}
\hline
               &\multicolumn{2}{c}{Reference$^{\rm a}$}   \\
\hline
 Object        &Optical & Infrared   \\	 
\hline
 \multicolumn{3}{c}{ELGs}   \\
 \noalign{\smallskip}
I Zw 18        &     15 &    38 \\
Haro 11        &     18 &    38 \\
UM 420	       &     13 &    38 \\
Mrk 1450       &     16 &    38 \\
NGC 4861       &     15 &    38 \\
UM 448	       &     13 &    38 \\
Mrk 930        &     13 &    38 \\
Mrk 162        &     13 &    38 \\
NGC1140        &     12 &    38 \\
UM461	       &     13 &    39 \\
UM462	       &     13 &    38 \\
Tol2138-405    &     21 &    38 \\
IIZw 40        &     19 &    38 \\
SBS0335-052E   &     14 &    38 \\
\hline
 \multicolumn{3}{c}{M\,101 \ion{H}{ii}Rs}   \\
 \noalign{\smallskip}
NGC5455        &     37 &     40 \\
NGC5461        &     37 &     40 \\
NGC5471        &     37 &     40 \\
\hline
 \multicolumn{3}{c}{ Magellanic \ion{H}{ii}Rs}   \\
 \noalign{\smallskip}
N160A1         &     41    & 41 \\
N160A2         &     41	   & 41 \\
N157B          &     41    & 41 \\
N4A            &     41    & 41 \\
N66            &     41    & 41 \\
N81            &     41    & 41 \\
\hline
\end{tabular}
\begin{minipage}[c]{1\columnwidth}
$^{\rm a}$ The reference number is according to the ID listed in Table~\ref{tab1}.
\end{minipage}
 \end{table}

\subsubsection{Instrumentation and observational techniques}

The data compiled from the literature were acquired using different
instrumentation and observational techniques. 
This may yield an additional scatter in the abundance determinations computed
along this work that might lead to biased results.
  For example,  if  emission-lines with a broad component are present
in  low resolution spectra these could not be detected,  and several 
physical properties obtained from these lines will be unreliable, such as
abundance determinations  \cite[see][and references therein]{hagele13}. 
Also, heavy elements abundance determinations from
spectra with low signal-to-noise ratio 
can be overestimated \citep{wesson12}.
 However, discrepancies in emission-line ratios estimated for a given object
observed using different techniques ranging from about 0.03 to 0.1\,dex
\citep{dors05} translate into a few percent in the uncertainties of ionic
abundance determinations.  
For example, if this discrepancy in the line ratio
[\ion{O}{iii}]($\lambda$5007+$\lambda$4959)/$\lambda$4363 is 0.06\,dex, the
electron temperature derived using the different observations differs by about
400\,K, yielding a log(Ne/O) variation of about 0.1\,dex. This value is of
the order of the intrinsic scatter found for uniform samples studied using
data obtained with the same instrumentation \cite[see e.g.][]{kennicutt03}.
When no systematic effects in the calibrations and data reduction exist, the
physical properties derived using data obtained with different observational techniques
should be the same. For example, the physical parameters estimated for the
star forming knots of Haro\,15 are in very good agreement 
when derived using long-slit and echelle data \citep{hagele12,lopez09}. 

\subsubsection{Nature of the sources}

Our sample is composed by objects with different morphology and sizes,
nevertheless they all have young and massive stellar clusters as their main
ionization sources.   
The shape of the emission-line spectra of the objects in our sample is
dominated by the ionizing fluxes from these massive stars.
Therefore the same measurement and analysis techniques can be applied to derive
the temperatures, densities and chemical composition of their interstellar gas
\citep{hagele06}.

In general, the \ion{H}{ii}Rs and ELGs are located in the same zone in
diagnostic diagrams \cite[see e.g.][]{bpt81} 
implying that similar ionization mechanisms
are taking place in these two different kind of objects (see Fig.\ \ref{fdisc1}).
Other ionizing  
mechanisms could be present according to the evolutionary stages and the
nature of the objects, such as galactic winds
due to supernovae \cite[e.g.][]{tang07}, contribution to the hydrogen emission
fluxes by diffuse ionized gas \cite[e.g.][]{oey07,moustakas06,calzetti04},
and/or the presence of obscured AGNs \citep{kewley01}, or even the effects of
the presence of multiple kinematical components in the emission-line profiles
\cite[see e.g.][and references therein]{hagele10,firpo10,firpo11,amorin12}.
The statistical contribution of these mechanisms is similar for the different
kind of objects and they do not yield any bias on the physical conditions of
the objects obtained in the present work.

\section{Photoionization models}
 \label{mod}
 
\begin{figure*}
\centering
\includegraphics[angle=-90,width=0.9\columnwidth]{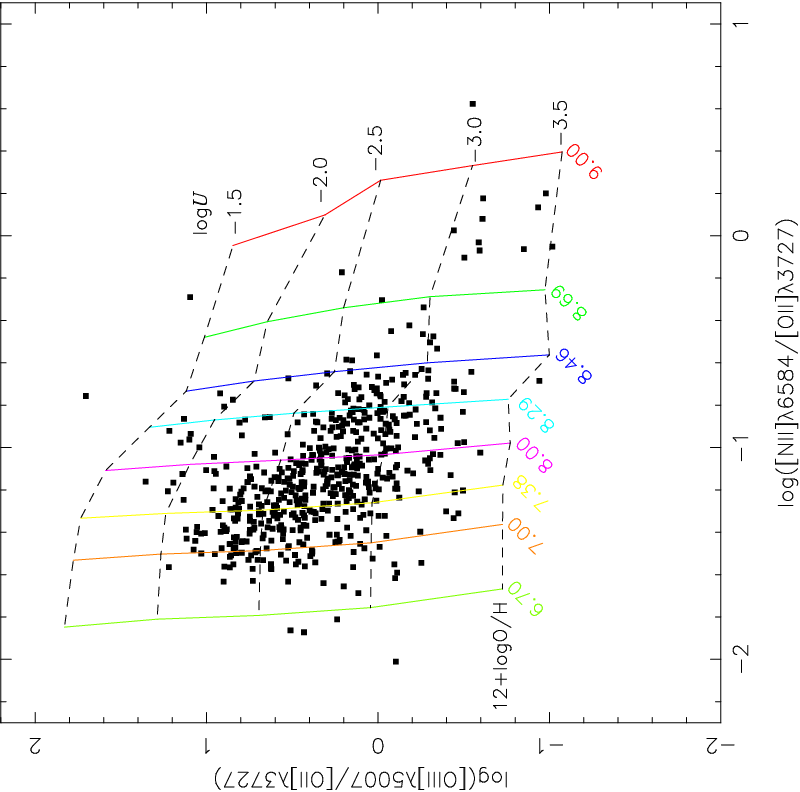}\hspace{0.4cm}
\includegraphics[angle=-90,width=0.9\columnwidth]{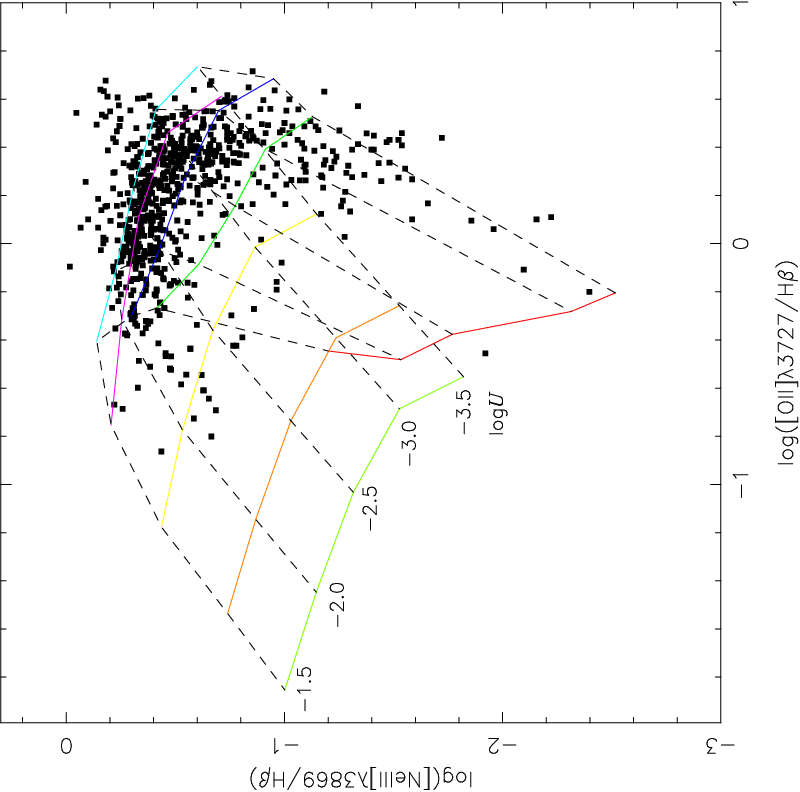}\\\vspace{0.4cm}
\includegraphics[angle=-90,width=0.9\columnwidth]{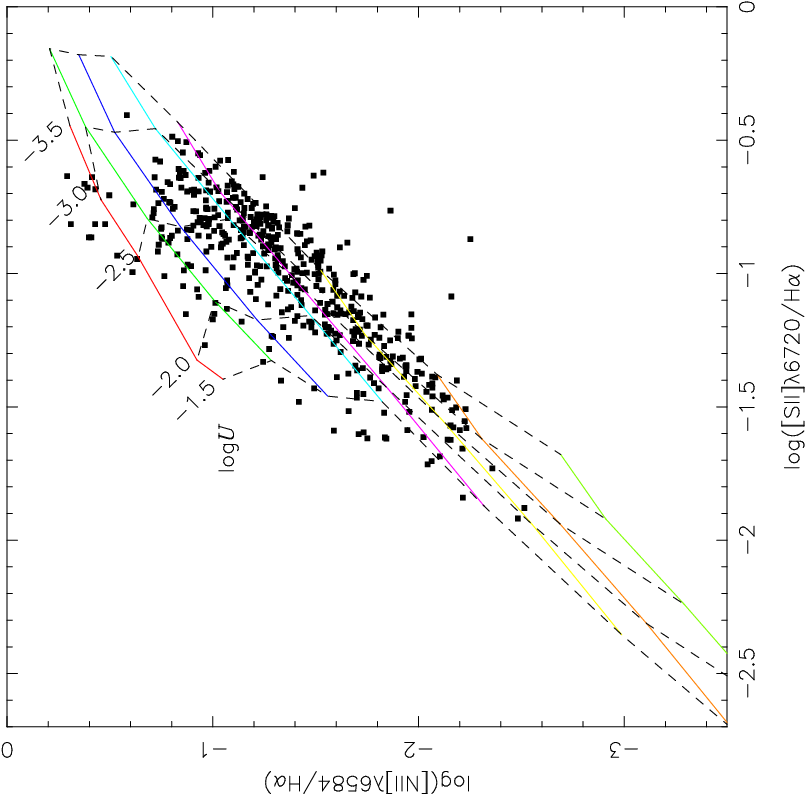}\hspace{0.4cm}
\includegraphics[angle=-90,width=0.9\columnwidth]{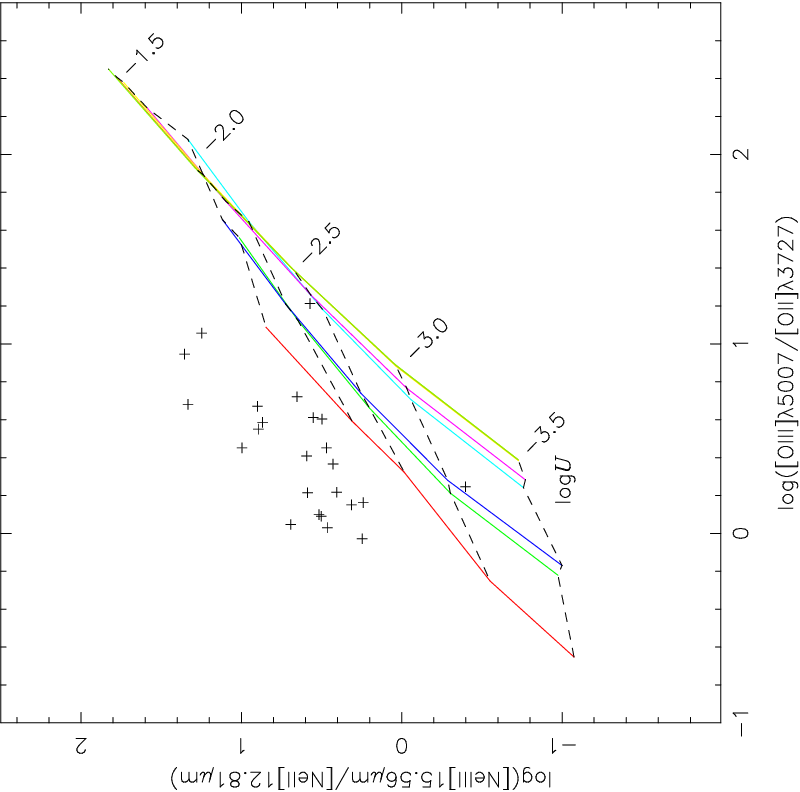}
\caption{ Diagnostic diagrams containing observational data taken from the
 literature (all groups) and the results of the grid of
 photoionization models described in this work. The solid lines connect
 curves of iso-(O/H), while the dashed lines connect curves of iso-$U$. 
 The colour code for the values of the oxygen abundances (tracer of the metallicity) are the
 same in all panels and are indicated in the upper left  panel.
  The solar  oxygen abundance  
refers to \citet{allende01}  and corresponds
to 12+log(O/H)=8.69. 
  The values of $\log U$ are indicated in  each panel.  Points are the
 observational data. The typical errors (bars not shown) of the observational emission-line
 ratio are about 10 per cent.} 
\label{f2}
\end{figure*}

 To estimate the ICF for Ne via photoionization models, we compared the results
of a grid of models performed using the Cloudy/8.0 code
\citep{ferland98} with the observational data sample. 
 We used data of the groups A, B, C,  D, and E
which include all objects with optical and IR emission-lines available (i.e. 734 objects).
The diagnostic diagrams
[\ion{O}{iii}]$\lambda$5007/[\ion{O}{ii}]$\lambda$3727
vs.\ [\ion{N}{ii}]$\lambda$6584/[\ion{O}{ii}]$\lambda$3727, 
[\ion{Ne}{iii}]$\lambda$3869/H$\beta$
vs.\ [\ion{O}{ii}]$\lambda$3727/H$\beta$, \newline
[\ion{N}{ii}]$\lambda$6584/H$\alpha$
vs.\ [\ion{S}{ii}]$\lambda$6720/H$\alpha$, and 
[\ion{Ne}{iii}]15.56$\mu$/[\ion{Ne}{ii}]12.81$\mu$m
vs.\ [\ion{O}{iii}]$\lambda$5007/[\ion{O}{ii}]$\lambda$3727 
were used for the comparison (see Fig.\ \ref{f2}). The ICF values were obtained from the model
results which better reproduce the observational data. 

The model grid was built following the same procedures as  \citet{dors11},
with metallicities of $Z=2.0, 1.0, 0.6, 0.4, 0.02, 0.01\: Z_{\odot}$, 
and logarithm of the ionization parameter $U$ from $-$3.5 to $-$1.5 dex with a step of $-$0.5 dex.
For each model, the ionizing source was assumed to be a stellar cluster
with a spectral energy distribution  obtained from the stellar
population synthesis code $Starburst99$ 
\citep{leitherer99}, with an upper mass limit of 100\,$M_{\odot}$ and an age of 1\,Myr. 
 For the hypothetical nebula we adopted a constant electron density $N_{\rm e}$= 200
 $\rm cm^{-3}$, plane-parallel geometry, and solar 
abundance ratio log(Ne/O)=$-0.61$ (oxygen from \citealt{allende01} and Ne from
\citealt{grevesse98}). 
In the models with $Z=2\:Z_{\odot}$, a value of log(Ne/O)=$-1.0$ was
assumed, in order to fit the data in this high metallicity regime. 
 This value is about the same found by \citet{perez07} for very high metallicities.
A complete description of the photoionization models is presented in \citet{dors11}. 

In Fig.~\ref{f2} the diagnostic diagrams containing the observed
emission-line intensities and the photoionization model results
are shown. The typical errors  of the observational emission-line
ratios are about 10 per cent.
It can be seen that the majority of the observational data falls within
the regions occupied by the models when all the considered emission lines are
in the optical range (see both upper and lower left panels). 
However, in lower right panel of Fig.~\ref{f2}, can be seen that the models 
produce lower values of the intensity line-ratio
[\ion{Ne}{iii}]15.56$\mu$m/[\ion{Ne}{ii}]12.81$\mu$m.
This problem was pointed out by \citet{perez09a} and by \citet{morisset04}.
They found similar results from photoionization model grids built assuming
different atmosphere models as photoionization sources. They concluded
that this effect could not be due to uncertainties in stellar atmosphere
models and   must  be due to problems in the atomic data of the mid-IR lines.

\section{Determination of ionic abundances}
\label{ion}

We considered two methods to determine the $\rm Ne^{++}$, $\rm Ne^{+}$, $\rm O^{+}$, and $\rm O^{++}$
ionic abundances in our sample. The
method that uses direct estimations of the electron temperatures
(this method will be called the Visible-lines method) and the one that uses infrared emission-lines 
(this method will be called the   IR-lines method). 
 
 \subsection{ Visible-lines method}
 \label{te}
 
 For those objects with the appropriate optical emission-line measurements,
  we calculated the 
electron temperature ($T_{\rm e}$) from the observed line-intensity ratio 
$R_{\rm O3}$=[\ion{O}{iii}]($\lambda4959\: + \: \lambda 5007)/\lambda4363$
for the high ionization zone (refereed as $t_{3}$). 
To do that we used the fitting function derived
by \citet{hagele08} 
based on the {\sc temden} routine of the nebular package of the {\sc iraf}\footnote{Image Reduction and Analysis Facility, distributed by NOAO, operated by AURA, Inc., under
agreement with NSF.} 
software:
 \begin{equation}
 \label{eqt3}
 t_{3}=0.8254-0.0002415 R_{\rm O3}+\frac{47.77}{R_{\rm O3}},
 \end{equation}
with $t$ in units of $10^{4}$K. The electron density ($N_{\rm e}$) was computed from the
ratio of [\ion{S}{ii}]$\lambda 6716/\lambda 6731$ using the {\sc temden}
routine and $t_{3}$ values given by the expression above.  

The $\rm O^{++}$ and $\rm Ne^{++}$ abundances were computed using $t_{3}$ and
following the relations given by \citet{perez07}: 
\begin{eqnarray}
 12+\log(\frac{{\rm O^{++}}}{{\rm H^{+}}}) \!\!\!&=&\!\!\! \log \big( \frac{I(4959)+I(5007)}{I{\rm (H\beta)}}\big)+6.144  \nonumber\\
                                          &&\!\!\!+\frac{1.251}{t_{3}}-0.55\log  t_{3}   
   \end{eqnarray}
   and
\begin{eqnarray}
\label{eq3a}
  12+\log(\frac{{\rm Ne^{++}}}{{\rm H^{+}}})  \!\!\!&=&\!\!\! \log  \big( \frac{I(3869)}{I{\rm (H\beta)}}\big)+6.486 \nonumber\\
                                           &&\!\!\!+\frac{1.558}{t_{3}}-0.504\log t_{3}.
\end{eqnarray}

The $\rm O^{+}$ ionic abundances were computed assuming $N_{\rm e}$\,=\,100\,cm$^{-3}$ and an electron
temperature of the low ionization zone (refereed as $t_{2}$) derived from
the theoretical relation: 
\begin{eqnarray}
\label{eqo3}
t_{2}^{-1}\,=\,0.693\,t_{3}^{-1}+0.281
\end{eqnarray}
obtained from the models described by \citet{perez03}.
Measurements of the [\ion{O}{ii}]$\lambda$7325 auroral emission line, sensitive
to the [\ion{O}{ii}]  electron temperature, are available for only  10-15
percent of our sample.
Since we used the observational sample
in a statistical way, we preferred to derive all the temperatures following
the same procedure. Besides, almost all objects present high or moderate excitation 
degrees   [($I$[\ion{O}{iii}]$\lambda$4959+$\lambda$5007)/($I$[\ion{O}{ii}]$\lambda$3727+
$I$[\ion{O}{iii}]$\lambda$4959+$\lambda$5007)\,$\geq$\,0.4, as was defined by
\citealt{pilyugin00}]. 
Therefore the impact of O$^+$/H$^+$ variations to the
total oxygen abundance and to the O$^{++}$/(O$^{++}$+O$^+$) ratio (used in 
the neon ICF determinations) is small   ($\approx$18\%)  in any case, and from a statistical
point of view negligible.
Only in those cases where the  [\ion{O}{iii}]$\lambda$4363 and  [\ion{O}{ii}]$\lambda$7325 auroral
emission-lines were not available, we computed the $\rm O^{+}$ ionic abundances
assuming the electron temperature directly estimated from the observed
line-intensity ratio $R_{\rm N2}$=[\ion{N}{ii}]($\lambda6548\: + \: \lambda
6584)/\lambda5755$: 
 \begin{equation}
 \label{eqn2}
 t_{2}=0.537-0.000253 R_{\rm N2}+\frac{42.13}{R_{\rm N2}}
 \end{equation}
fitting function given by \citet{hagele08}.    For few objects ($< 1$\%)  
$t_{2}$ was derived using this equation.

To derive the $\rm O^{+}$ ionic abundances we used the expression given by
H\"agele and collaborators:
\begin{eqnarray}
 12+\log(\frac{{\rm O^{+}}}{{\rm H^{+}}})  \!\!\!&=&\!\!\! \log  \big( \frac{I(3727)}{I{\rm (H\beta)}}\big)+5.992 \nonumber\\
                                           &&\!\!\!+\frac{1.583}{t_{2}}-0.681\log t_{2} +\log(1+2.3 n_{\rm e}),
  \end{eqnarray}

\noindent where $n_{\rm e}=N_{\rm e}/(10^{4} \rm cm^{-3})$. As pointed out by
these authors, the expressions
given above are valid for temperatures from 7\,000 to 23\,000\,K. 
  The  $\rm O^{+}$ ionic abundances  calculated using    $t_{2}$ values  from Eqs.~\ref{eqo3} and \ref{eqn2}
   differ by about 0.2 dex. 
The data set used to determine the ionic abundances using this method belongs
to groups A and B and is comprised by 579 objects ($\sim80$\% of the sample).
All $\rm O^{+}$, $\rm O^{++}$ and $\rm Ne^{++}$ ionic abundances obtained via Visible method used along
this study were computed from this data set.
 
\subsection{  IR-lines method}
\label{corr}

 The $\rm Ne^{+}$ and $\rm Ne^{++}$ ionic fractions can be determined
  using the intensities of the [\ion{Ne}{ii}]12.81$\mu$m and
  [\ion{Ne}{iii}]15.56$\mu$m emission-lines, respectively, and following a
  similar methodology 
to the one presented by \citet{forster01}. The ionic
abundance of an element can be given by 

 \begin{equation}
 \label{eq1}
 \frac{n_{  {\rm X^{+i}}}}{ n_{{\rm H^{+}}}}=\frac{I_{\lambda}( {\rm X^{+i})}  \: N_{{\rm e} }        \: j_{\lambda( {\rm  H^{+}})}}{I_{\lambda}( {\rm  H^{+}) } \: N_{{\rm e} } \:
 j_{\lambda(  {\rm X^{+i}})} },  
   \end{equation}
  where  $n_{  {\rm X^{+i}}}$  and  $n_{{\rm H^{+}}}$
 are the densities  of the ions  $\rm {\rm X^{+i}}$  and   $\rm H^{+}$,  $I_{\lambda}( {\rm X^{+i})}$ is intensity of a given emission-line emitted by $\rm {\rm X^{+i}}$, 
 $I_{\lambda(\rm H^{+})}$ is the  intensity of a reference hydrogen line,
 while   $j_{\lambda(  {\rm X^{+i}}   )}$ and    $j _{\lambda( \rm  H^{+} )}$
 are the emissivity values given by the {\sc ionic} routine of the nebular
 package of the {\sc iraf} software which uses the Ne atomic parameters  from 
\citet{badnell06}, \citet{griffin01}, \citet{kaufman86}, \citet{butler94}, and
\citet{mendoza83}. These emissivity values are assumed to be constant in all
abundance determinations because these vary less than 5\% over a large
temperature range   (see the pioneer work by \citealt{simpson75}).

Using this method any error in the determination of these emissivities
directly translates in a systematic shift in the derived Ne$^{+}$ and
Ne$^{++}$ ionic abundances. Taking into account all these assumptions we
obtained
 \begin{equation}
 \label{eq11}
 \frac{{\rm Ne^{+}}}{{\rm H^{+}}}=\frac{I(12.81\mu {\rm m})}{I{\rm (H\beta)}} \: \times \: 1.322\:10^{-4},
  \end{equation}
 and
 \begin{equation}
 \label{eq12}
 \frac{{\rm Ne^{++}}}{{\rm H^{+}}}=\frac{I(15.56\mu {\rm m})}{I{\rm (H\beta)}} \: \times \: 6.323\:10^{-5}.
  \end{equation}

\section{Comparison of neon ionic determinations}
\label{compneon}

\begin{figure}
\centering
\includegraphics[angle=-90,width=8cm]{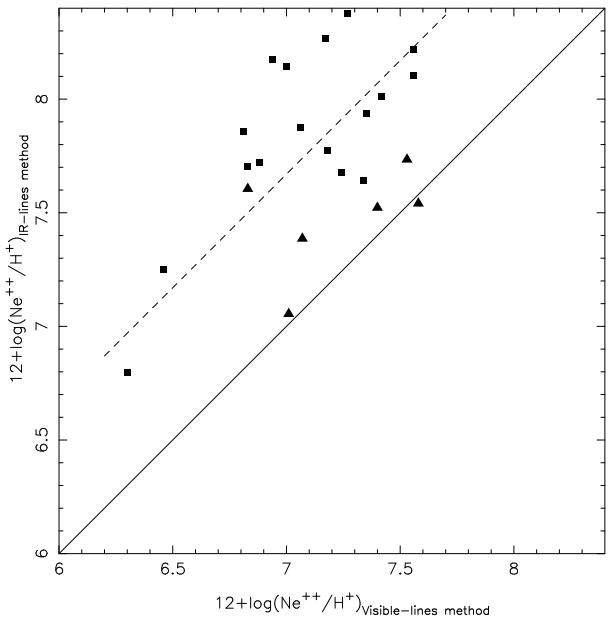}
\caption{Comparison between ionic abundance of the $\rm Ne^{++}/H^{+}$
 derived using  Visible-lines and IR-lines methods. Points represent estimations for the objects presented in Table~\ref{tab1a},   where
 squares are estimations using H$\beta$ fluxes  and triangles the ones using  Br$\beta$ fluxes. 
 Solid line represents equality
 of the two estimates. Dashed line is the equality line shifted by the
 average of the differences between the ionic abundances. Typical errors in
 optical and infrared estimations are of about 15\% and 10\%, respectively \citep{vermeij02a}.}
\label{f0}
\end{figure}

 With the aim to compare the $\rm Ne^{++}/H^{+}$ ionic abundances derived using the
   Visible-lines method and those using the  IR-lines method, we
  plotted in Fig.\ \ref{f0} the results for the 23 objects listed in
  Table~\ref{tab1a}, the only ones for which we are able to apply both
  methods. These objects are those that simultaneously belong to groups A or B, and
  E. We can note a large discrepancy and scatter. Comparing the $\rm Ne^{++}/H^{+} $ ionic abundance determinations obtained via
the  Visible-lines method  with those based on the   IR-lines method, we
found discrepancies of about a factor of 4 in average, being 
$\rm Ne^{++}/H^{+}$  underestimated when using the Visible-lines method
(see Fig.\ \ref{f0}).

 \begin{figure}
\centering
\includegraphics[angle=-90,width=1\columnwidth]{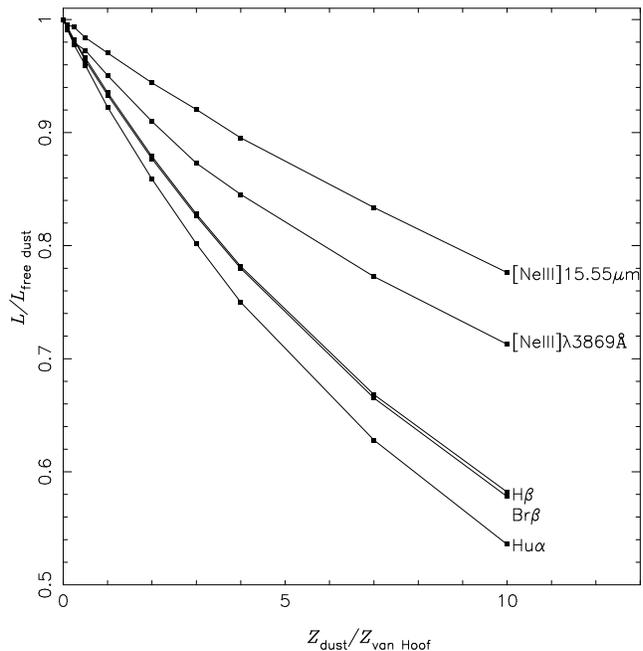}
\caption{Logarithm of  the  luminosity predicted by our models  for some hydrogen and neon emission-lines versus
 the dust internal  abundance   in relation to the  default value   assumed in the
 Cloudy code,  i.e.  $\rm \log\:O/H\approx -4$. The grain model is described in \citet{hoof01}.
 Points are results for models considering
 different dust abundances connect by the curves.    The results for different emission-lines are indicated.
 The $Z_{\rm dust}/Z_{\rm van \:Hoof}$=0 represents the models free of internal dust.} 
\label{fdust}
\end{figure}

\begin{figure}
\centering
\includegraphics[angle=-90,width=0.95\columnwidth]{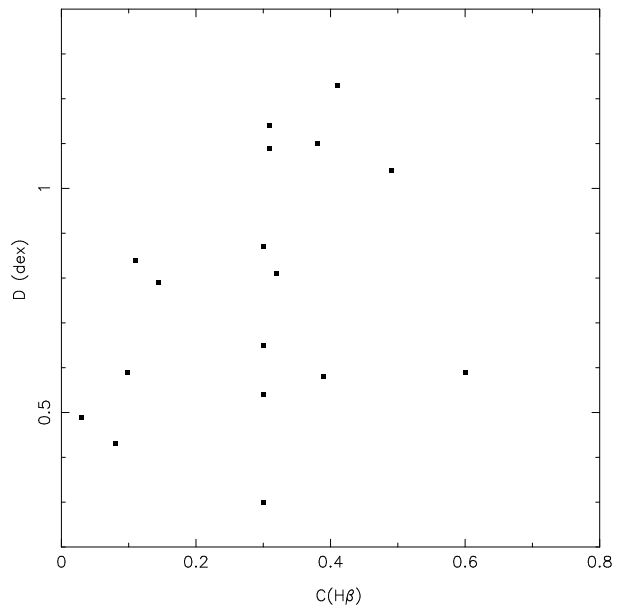}
\caption{Differences between Ne$^{++}$ ionic abundances obtained via the IR-lines
  and the Visible-lines methods as a function of the nebular reddening constant
C(H$\beta$) for the majority of the objects listed in Table~\ref{tab1a}.} 
\label{f55}
\end{figure}

The origin of this discrepancy is uncertain and several explanations have been proposed. 
\citet{vermeij02a} compared optical and infrared
ionic fractions obtained for a sample of \ion{H}{ii}Rs located in the Magellanic Clouds.
These authors found considerable discrepancies between determinations via these two methods, which
were attributed to the difference in the area of the sky covered by the infrared and
optical observations. However, as discussed above, aperture effects are not 
probably the origin of 
this huge discrepancy regarding a high-excitation ion as Ne$^{++}$.

Extinction might noticeably affect the comparison between
the abundances obtained from IR and optical emission-lines. The blue optical
[\ion{Ne}{iii}] emission-line is more absorbed by dust than the mid-IR
ones. Hence, if the nebular emission is highly reddened by dust the optical
line does not trace all the Ne in the inner parts of the SFRs, yielding an
optical ionic abundance lower than that derived from the mid-IR lines.
In the same way as for the Ne emission-lines, this effect could be enhanced
by the use of H$\beta$ or Br$\beta$ emission 
lines instead of Hu$\alpha$ to derive ionic abundances in the mid-IR. 
  In order to verify the extinction influence on the   emission-lines, 
 models with different internal dust abundance were built.
 In these, the only effect of
grains on the  continuous opacity was considered, i.e. the gas heating by grain emission
was not considered  in the calculations.    The dust abundance was linearly scaled with the 
default value  assumed in the
 Cloudy code,  i.e.  $\rm \log\:O/H\approx -4$.  The grain model is described in \citet{hoof01}.
In Fig.~\ref{fdust} the predicted luminosity values  ($L/L_{\rm{free\,dust}}$)
for some hydrogen and neon emission-lines in relation to the ones predicted by the model 
free of dust ($Z_{\rm dust}/Z_{\rm van \:Hoof}$=0)  are shown. We can see  that   
H$\beta$, Br$\beta$  and Hu$\alpha$ have about the same decreasing with the
dust abundance,    reaching $\sim$4\% maximum difference in their
luminosities for $Z_{\rm dust}/Z_{\rm van \:Hoof}$=10.  A
similar behaviour is found for the neon emission-lines with a difference of
about 6\%. Even though these differences are within the observational
measurement errors,
we have  used these predicted emission-line intensities to calculate the $\rm
Ne^{++}$ 
abundance following the same procedure  presented in  Section~\ref{ion}.
 The largest difference  found in  $\rm Ne^{++}$ abundances derived  from IR
 and the Visible  methods was  
 0.3 dex. Therefore,  this result shows that, at least, the  extinction caused
 by the internal dust  
is not  the main source of the neon ionic discrepancy.

To verify if there is an observational correlation between the extinction and the abundance
discrepancy for the 23 objects listed in Table \ref{tab1a}, we plotted in
Fig.~\ref{f55} the difference (D) between the  
Ne$^{++}$ ionic abundances obtained via the IR-lines and the Visible-lines methods  
versus the nebular reddening constant 
[C(H$\beta$)] compiled from the literature and used by the authors to
correct their observational data. 
There is no evident correlation between the discrepancy and the extinction
found from the Balmer decrement.  

  For  five objects of our sample: N81, N4A, N66, N160A1, and N160A2, 
both Br$\beta$  and  H$\beta$ fluxes   were obtained  directly from
spectroscopy observations by \citet{vermeij02}, \citet{dufour77a, dufour77b} and \citet{heydari88, heydari02}.
We use these data and the [\ion{Ne}{iii}]15.55$\mu$m from  \citet{vermeij02} to calculate 
 $\rm Ne^{++}/H^{+}$  ionic  abundances of these objects using both hydrogen lines. We found 
 neon ionic estimations in  agreement  by about 0.2 dex, indicating that dust absorption is not the main source of
 the discrepancy,  at least for these objects.  Another observational
 support  to this result arises from the data presented by  \citet{wu08}.
These authors pointed out that  most of the  galaxies observed by them, which include 50\%  of the  ELGs presented in Table~\ref{tab1a},
do not show a strong 9.7$\mu$m silicate feature and, consequently, have low dust extinction.
  Moreover, as noted in Fig.~\ref{f0}, for the objects  whose both Br$\beta$  and  H$\beta$ fluxes were measured,
the ionic difference is also found, indicating that the differential absorption of optical and infrared
emission lines is not the main cause of the discrepancy. Thus, we can assume that this
effect is negligible for these objects.

As can be seen in Sect.~\ref{mod}, and was already noticed by \citet{perez09a} and 
\citet{morisset04}, the photoionization models
can not reproduce diagnostic diagrams based on Ne mid-IR lines independently of
the used stellar atmospheres. Therefore,  this could be symptomatic of inaccurate 
atomic data involving these emission-lines and can be on the basis of the 
found discrepancies when deriving abundances from optical and mid-IR emission
lines. 
 In particular, the atomic parameter which can affect the abundance determinations is
the collision strength for the [\ion{Ne}{iii}]$\lambda15.56\mu{\rm m}$
emission-line. However, it is not probable
that uncertainties in this parameter be the main cause of the discrepancy
found in our study.
Along the years differences in neon collision strength have been found. For example,
the collision strength for [\ion{Ne}{iii}] $^{3}\!P_{1}\rightarrow ^{3}\!\!\!\!P_{2}$ derived by \citet{blaha69} is 
$\Omega$=0.581. The $\Omega$ values derived in the IRON project by \citet{butler94} 
range from 0.481 to 0.778 for temperatures ranging from $10^{3}$ to
$10^{5}$\,K. Later computations by \citet{mclaughlin00} yield a lower $\Omega$ value
by about a 20\% than the one proposed by \citet{blaha69}. Recently, \citet{mclaughlin11} have recomputed the collision strengths for some neon lines using
a small scale 56-level Breit-Pauli calculation and a large-scale 554-levels R-matrix Intermediate Coupling Frame Transformation (ICFT).
They showed that different $\Omega$ values are obtained by the use of these suppositions,  in the sense
 that for  56-level Breit-Pauli calculation  $\Omega$ ranges from about  0.5 to 1.0 and for the  554-levels ICFT
 ranges from about 0.4 to 0.8. It must be noted that these results are in good
agreement with earlier derivations.  Nevertheless, e  ven assuming  that
  the collision strength for the  [\ion{Ne}{iii}]$\lambda15.56\mu{\rm m}$ emission-line can vary by a factor of two, 
 it is not enough to conciliate the ionic abundance discrepancy.

Finally, another supposition is the presence of electron temperature
fluctuation amplitude inside the ionized
gas. It can lead to an underestimation of the ionic abundances obtained via
optical collisionally-excited lines when these fluctuations are not considered since these
lines are strongly dependent on the electron temperature
\cite[see e.g.][]{peimbert67,peimbert2003,krabbe05}. This problem can be alleviated using abundances based 
on IR emission-lines, whose fine-structure transitions have a weak dependence on the electron temperature
\cite[e.g.][]{dinerstein85,rubin01,nollemberg02,garnett04}.
Since the electron temperatures are related with the chemical composition of
the emitting gas, temperature fluctuations are essentially equivalent to 
abundance variations across the nebula \citep{kingdon98}.

To test the effect of the presence of temperature fluctuations and/or
abundance variations, we built a grid of photoionization models using Cloudy.
These models are similar to the ones described in Sect.~\ref{mod}  and presented in in Fig.~\ref{f2} but
considering the presence of chemical inhomogeneities in the theoretical
nebula  and  spherical geometry.  
  The   plane-parallel geometry assumed in the  later models can not be used  to test  variations along the  radius  of the
hypothetical nebulae, since makes the inner radius much larger than the thickness of the cloud and  the results
about independent on  electron temperature fluctuation amplitude.
In this case we considered a nebula with $Z=Z_{\odot}$,   with an inner
radius of 4\,pc and an outer radius where the temperature falls below 4000\,K
($\approx$30\,pc), and a number of hydrogen-ionizing photons emitted by the
ionizing star cluster of $\rm 10^{50} \:s^{-1}$. 
It was assumed that the metallicity varies along the radius of the nebula as
a sine wave, with period P ranging from 0.01   to 2\,pc, including
0.01, 0.1, 0.25, 0.5 and 1 pc.   Also a model with  no variations and solar abundance was considered.
 We considered a set of models with an abundance amplitude of 0.8\,dex, which
takes metallicity values between $2\times Z_{\odot}$ and $0.3\times Z_{\odot}$.
\citet{kingdon95} built grids of photoionization models using similar sine
wave variations but applied to the total hydrogen density across the nebula.

\begin{figure}
\centering
\includegraphics[angle=-90,width=1\columnwidth]{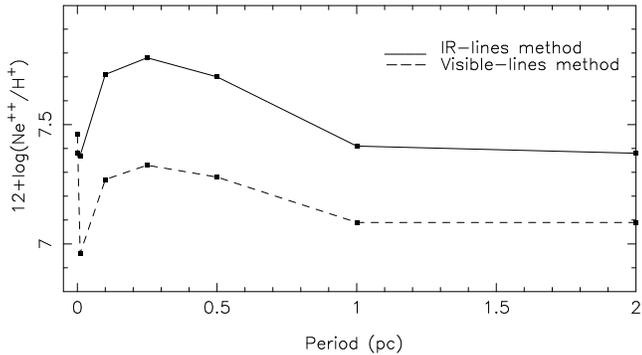}
\caption{Results of photoionization models considering abundance variations
along the radius of the nebula. $\rm Ne^{++}/H^{+}$ ionic abundances obtained
via   the Visible-lines  and  the IR-lines methods (dashed- and
solid-lines, respectively) as a function 
of the period P of the abundance variations.}
\label{f10}
\end{figure}

\begin{figure}
\centering
\includegraphics[angle=-90,width=1\columnwidth]{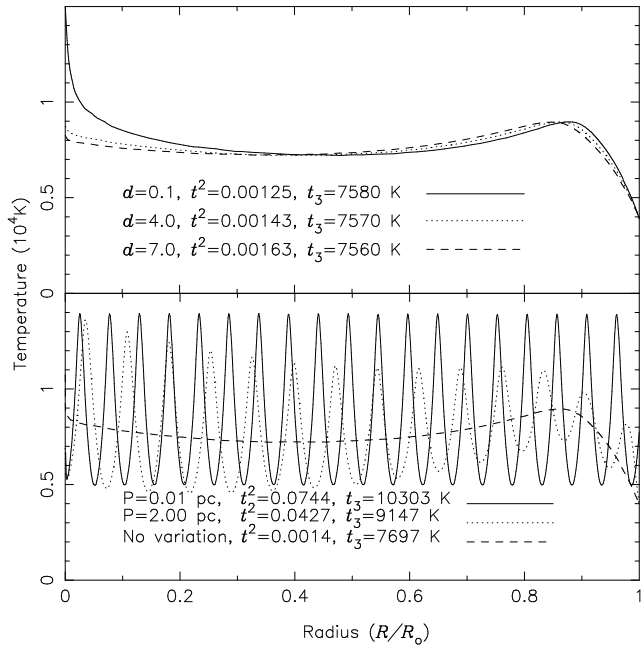}
\caption{Temperature structure of the theoretical nebula from our
  photoionization models. For each model, $R/R_{\rm o}$ is the radius normalized to the
  outermost one ($R_{\rm o}$). Bottom panel: The curves represent result models with different  values of 
  P (period of the sine wave which the  metallicity varies along the radius of the nebula) as indicated. 
   Top Panel: The curves represent model results with   no abundance variations
  but with different distances  between the central ionizing source and the inner face of the gas, represented by $d$, as indicated. The  $t^{2}$   
  as defined by \citet{peimbert67}  and   the electron temperature $t{_3}$  for each models are  shown.} 
\label{estru}
\end{figure}

  We used the emission-line intensities given by our models for each value
of P to calculate the $\rm Ne^{++}/H^{+}$ ionic abundances applying the
Visible-lines  and the IR-lines methods following the same procedure as
described in Section~\ref{ion}. Thus, we obtain two estimations of $\rm
Ne^{++}/H^{+}$ for each model.   
Fig.~\ref{f10} shows these estimated values of $\rm Ne^{++}/H^{+}$ plotted as
a function of the period P of the abundance variations. 
As it can be seen when
no variations are considered, the $\rm Ne^{++}/H^{+}$ ionic abundances
estimated by each method differ by less than 0.1 dex,   which is within the errors of
the methods.
When abundance variations are present the differences
could reach about 0.5 dex, being the ionic abundances via the
 IR-lines method the 
largest one. This value is very similar to the one found in Fig.\ \ref{f0} between
the estimated Ne ionic abundances derived applying the   Visible-lines  and the
  IR-lines methods  to the observational data. Moreover, these differences, as 
expected, depend on the metallicity amplitude of the models, increasing
(decreasing) with an increment (decrement) in the amplitude  (models not shown).
Therefore, the  temperature fluctuations studied through the abundance variations across
the nebula could be a probable explanation for the discrepancy between the
estimated Ne abundances by these two methods.
 In Fig.~\ref{estru} (bottom panel), we have plotted the temperature structure \footnote{In  Cloudy code the nebula is divided   in concentric zones
 with  physical conditions across them  about constant. This  causes  the zone number in each computed model to vary. The electron temperatures of these
zones are shown in Fig.~\ref{estru}. See Cloudy manual for more details.} of our
models with P=2, 0.01 and  no fluctuations  versus the radius normalized to
the outermost one.  In this figure we also show, for each P, the values of the temperature
fluctuation parameter ($t^{2}$) as defined by \citet{peimbert67} 
  and  the electron temperature $t_{3}$ using the predicted values of $R_{\rm O3}$  and the Eq.~\ref{eqt3}.

We found, as expected, that abundance variations translates into
temperature fluctuations. As can be  seen in Fig.~\ref{estru},    $t_{3}$   decreases when larger values of P  
are assumed,  however, the  average temperature weighted by the electron density over the volume
of the nebula $<$T$>$ (not shown), as defined by \citet{peimbert67}, is about constant with P.
The model with P=0.01\,pc shows a constant temperature fluctuation amplitude of about
9000\,K, while the amplitude showed by the model with P=2 decreases with the
radius. Interestingly, the $t^{2}$ value, when no fluctuations are considered, is about the same
found in some \ion{H}{ii}Rs \cite[see e.g.][]{krabbe02} and \ion{H}{ii}Gs \cite[e.g.][]{hagele06}. 
On the other hand, the $t^{2}$ values that explain the differences found when
comparing ionic abundances obtained via recombination and collisional
excited emission-lines \cite[see e.g.][]{peimbert2003,garcia-rojas06} are
in good agreement with those given by the models with P different than zero.
 A more detailed study about  spatial electron temperature fluctuations in ionized nebulae 
from photoionization models is given by \citet{copetti06}.

 Along decades observations have failed in found a  direct evidence  of chemical inhomogeneities in \ion{H}{ii} regions
(e.g. \citealt{oliveira08, lebouteiller8, garcia-rojas06, krabbe02}) and in planetary nebulae (e.g. \citealt{krabbe05, rubin02, liu00}).
Thus, an another source of temperature fluctuations can be the presence of density variations in the gas.
\citet{rubin89} showed that  large-scale variations in 
electron density in the gas do  arise  electron temperature fluctuations.
Considering the \ion{H}{ii} region models of \citet{rubin85}, this author pointed out that if
$N_{\rm e}$ varies by a factor of about 10, the electron temperature
can range  up to 2000 K  (see also \citealt{kingdon95}).
Recently, \citet{mesa12} presented results from integral field spectroscopy of a region near to the Trapezium Cluster 
and found the presence of  high-density gas ($N_{\rm e}\approx 4 \times 10^{5}\: \rm cm^{-3}$), which are much higher than the typical values obtained 
either for galactic and extragalactic regions (see \citealt{copetti00}).  Unfortunately,   
a larger number of observations, such  as the ones by \citet{mesa12},  are needed to confirm the high scale density  fluctuations
in ionized nebulae.

Another likely reason for  the presence  temperature fluctuations are different proximities of gas to the ionizing O stars,
 since these are mainly responsible for the ionization structure. To test this, we built models considering different distances, $d$, between
the central ionizing source and the inner face of the gas, where  $d$   ranged from 0.1 to 7 pc. These values are consistent with the ones
used by \citet{simpson97} in a study of the Galactic \ion{H}{ii} region G0.18-0.04. Again, the predicted  emission-lines  were used to 
 calculate the $\rm Ne^{++}/H^{+}$ ionic abundances applying the  Visible-lines  and the  
  IR-lines methods, following the same procedure as described in
  Section~\ref{ion}.    We found a 0.0-0.2\,dex difference in $\rm Ne^{++}/H^{+}$.  In Fig.~\ref{estru} (top panel) the temperature structure  of the 
 models is shown.   We can note a hot region very near the central star, however,
 this region is too  small to affect the derived abundances, since no significant changes among the
 models  neither  in the $t^{2}$ nor in  $t_{3}$ values are obtained. 
  Models with a  more complicated geometry, with many
stars distributed through the gas  (e.g. \citealt{ercolano07}), might cause the temperature
variations and consequently help to explain $\rm Ne^{++}/H^{+}$ ionic discrepancies.

\section{Neon ICFs}
\label{sicf}

\subsection{ICFs determinations} 

The ICFs stand for the unseen ionization stages of each element.
For the $\rm Ne^{++}$ it is defined as 
 \begin{equation}
 \rm ICF(Ne^{++})=Ne/Ne^{++}.
 \label{icf-def}
 \end{equation}

\begin{figure}
\centering
\includegraphics[angle=0,width=1.00\columnwidth]{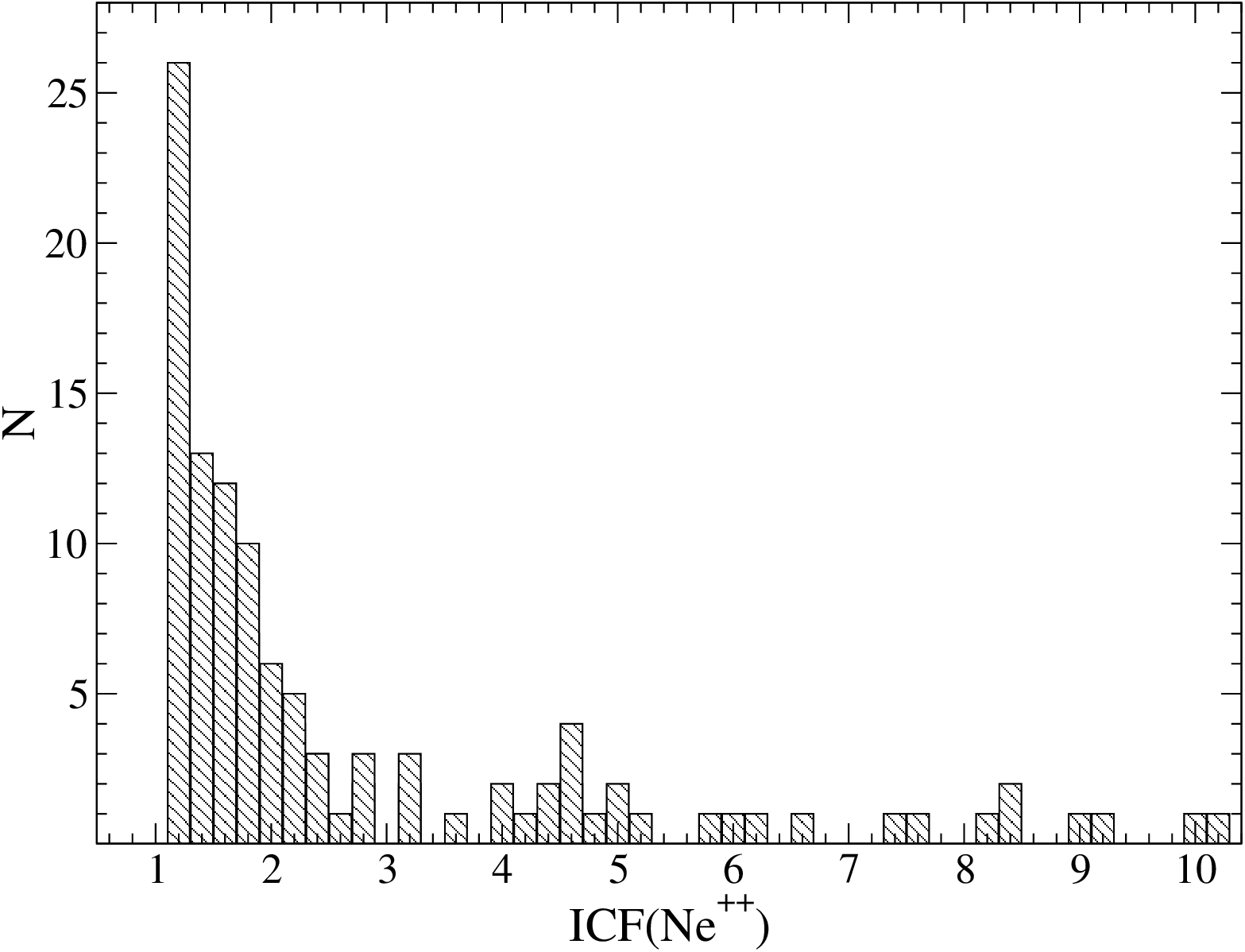}
\caption{Histogram containing the values of the ICF($\rm Ne^{++}$) obtained
 using the IR emission-line intensities and Eq.~\ref{eq3} for the
 objects belonging the group E and with ICF values lower than 10.}
\label{f6}
\end{figure}

The ICF for $\rm Ne^{++}$ can be directly computed using only 
 [\ion{Ne}{ii}]12.81$\mu$m and [\ion{Ne}{iii}]15.56$\mu$m emission-line fluxes.
 Assuming that the total Ne abundance is 
 \begin{equation}
  \frac{{\rm Ne}}{{\rm H}} \approx \frac{{\rm Ne^{+}}}{{\rm H^{+}}}+ \frac{{\rm Ne^{++}}}{{{\rm H^{+}}}}, 
 \end{equation}
 and using Eq.\ \ref{icf-def} we obtained
  \begin{equation} 
  \label{eq2}
   {\rm ICF(Ne^{++})}=1+\frac{ {\rm Ne^{+}}}{ {\rm Ne^{++}}}.
  \end{equation}
Using Eqs.\ \ref{eq11} and \ref{eq12}, we found  
   \begin{equation}
   \label{eq3}
  {\rm ICF(Ne^{++})}=1+2.10 \:  \times \: \frac{F_{[{\rm Ne\:II}]\lambda12.81\mu{\rm m}}}{F_{[{\rm Ne\:III}]\lambda15.56\mu{\rm m}}}. 
  \end{equation}
 Therefore to estimate the Ne ICF using the IR lines it is not necessary the use of
  hydrogen emission-line fluxes. 
  Thus, we were able to use all the objects belonging to group E (143
  objects).
A histogram containing the ICF values lower than 10 is shown in Fig.~\ref{f6}. 
It can be seen that almost all the ICF values   ($\approx 85$\%) are lower than 5. 
Very high values ($>$\,10) are also found. 
The objects that present the largest ICF values are 
IC342 ($\approx$\,257), NGC\,5236 and NGC\,5253 ($\approx$\,58), Searle\,5
($\approx$\,55), NGC\,3628 ($\approx$\,45), and NGC\,2903
($\approx$\,32).   Also high ICFs were obtained from data of some
\ion{H}{ii}Rs in M\,83 observed by \citet{rubin07}.   These values  (not shown)
could reflect the presence of cool ionizing sources in these objects.

\begin{figure}
\centering
\includegraphics[angle=-90,width=1.00\columnwidth]{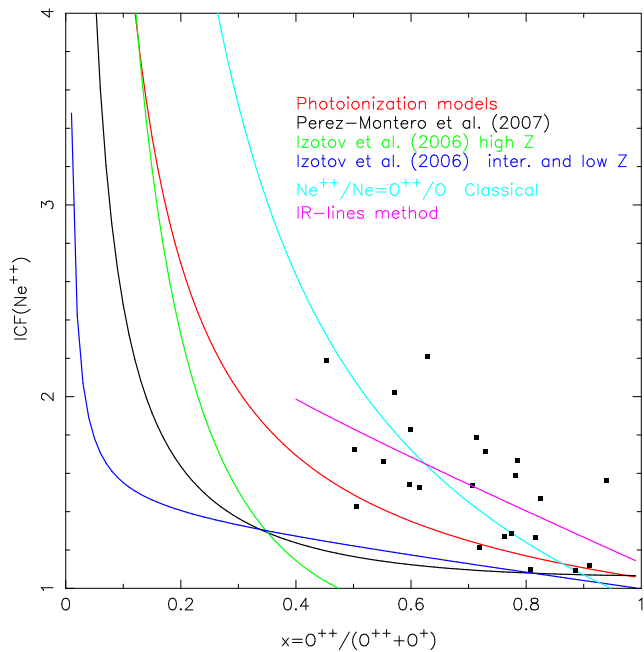}
\caption{Relation between ICF($\rm Ne^{++}$) and x=$\rm O^{++}/(O^{+}+O^{++})$
ionic abundance ratio. Solid lines correspond to our fittings to the ICF for Ne
and those found by different authors as indicated. Points represent the data
whose ICFs were determined via IR-lines method and listed in
Table~\ref{tab1a}.} 
\label{f5}
\end{figure}

 A classical way to determine the neon ICF takes advantage of its relation
  with the ionic oxygen $\rm O^{++}/(O^{+}+O^{++})$ ratio.
From a fit in our photoionized model results   presented in Sect.~\ref{mod} we obtained:
 \begin{equation}
  \label{eq4}
\rm ICF(Ne^{++})_{model}=0.741-0.08 x+\frac{0.393}{\rm x}, 
\end{equation}
where x=$\rm O^{++}/(O^{+}+O^{++})$. This relation is shown in Fig.~\ref{f5}.
In this figure are 
also shown the ICFs obtained using photoionization 
models by \citet[][]{perez07} and by \citet[][]{izotov06b}, the classical
approximation [ICF($\rm Ne^{++}$)=O/$\rm
O^{++}$]   which the  ionic abundances were derived from Visible-lines method and using our
data sample,  and the ICFs obtained from the IR-lines method. 
 
Fitting the results obtained using the  ICF via IR-lines method and x via Visible-lines method, represented by points in
Fig.~\ref{f5}, we obtained the following expression:
 \begin{equation}
  \label{eq5}
\rm ICF(Ne^{++})_{IR}=2.382-1.301x+\frac{0.05}{x}, 
\end{equation}
valid for $\rm x \: \geq \: 0.4$.
 Unfortunately, only for 23 objects in our sample we have measurements of
the Ne infrared emission-line intensities and we were able to determine the 
$\rm O^{++}/(O^{+}+O^{++})$ ratio via the Visible-lines method (see Fig.\ \ref{f5}).
It must be noted that this sample comprehends objects in a large range of metallicity 
($7.1 \: < \rm \: 12+log(O/H) \: < \:8.50$ dex) as well as a large range of
ionization degree ($0.4 \: < \rm O^{++}/(O^{+}+O^{++})  \: < \:1.0$).

\subsection{Application of the Neon ICFs}
\label{disc}

\begin{figure*}
\centering
\includegraphics[angle=-90,width=1.0\textwidth]{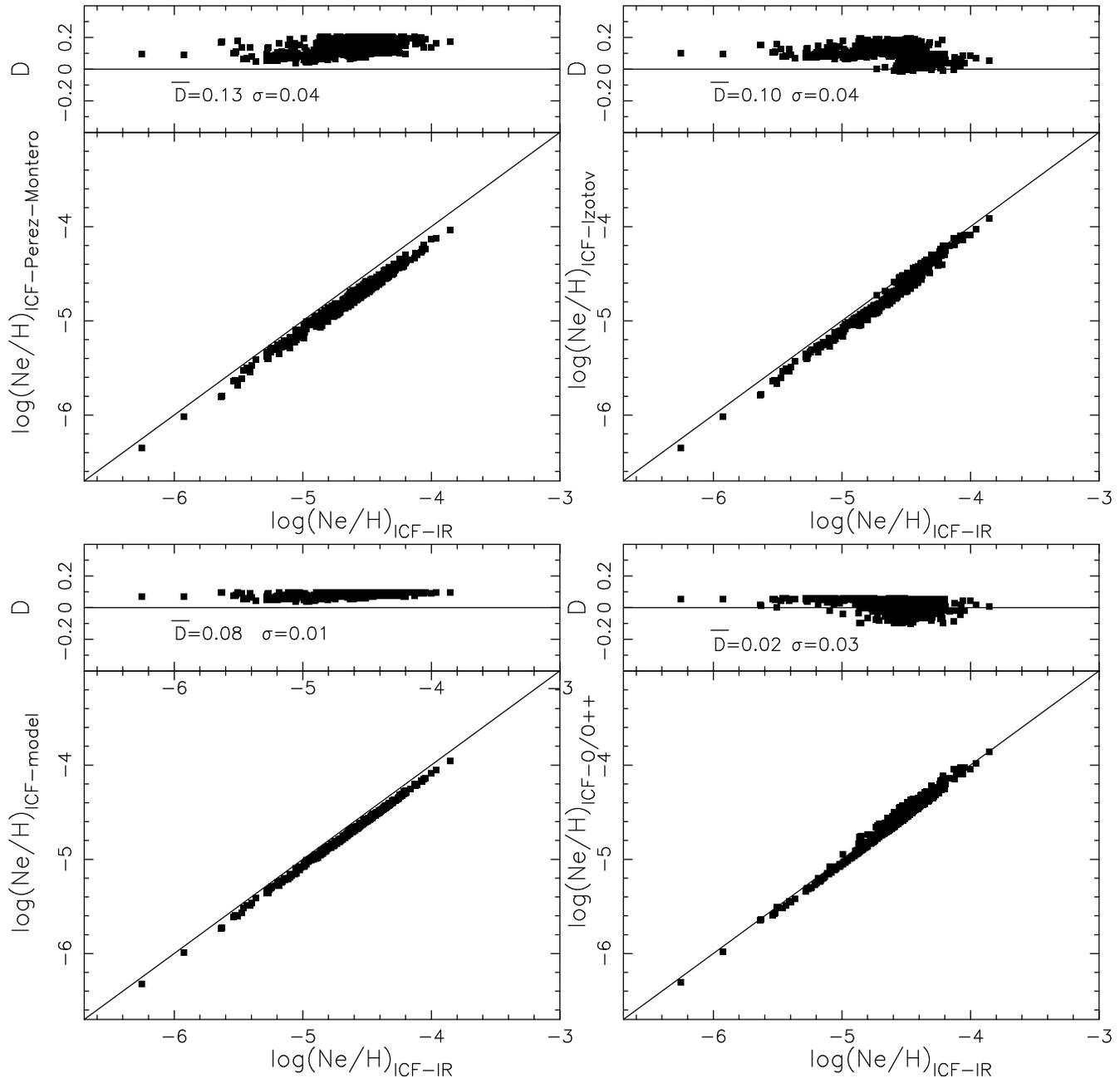}
\caption{ Comparison between total Ne/H abundances obtained from the 
    $\rm Ne^{++}/H^{+}$ ionic ones using the Visible-lines method and applying
    different ICFs as indicated. Points represent estimations for the objects
    belonging to the groups A and B.
 The top panel of each diagram shows the difference
 between the estimations using the methods considered. The average value
 $\overline{{\textrm D}}$ of this difference and the dispersion are shown
in each plot. Solid lines represent equality of the two estimates.}
\label{f0a}
\end{figure*}

  To test the differences in Ne/H total abundance yielded by the use of different ICFs,
we used the observational data to compute  Ne$^{++}$/H$^{+}$ ionic
abundances via
the Visible-lines method and applied different ICFs to obtain the Ne total abundances.
Fig.\ \ref{f0a} shows the comparisons between the Ne/H derived using the ICF from
the IR-lines method (Eq.~\ref{eq5}) and those values estimated applying the ICFs derived by:
(a) \citet{perez07}, (b) \citet{izotov06b}, (c) 
photoionization models (Eq.~\ref{eq4}) described in this work, and (d) the classical approximation [${\rm
 ICF(Ne^{++})}=\rm  O/O^{++}$]. 
It can be seen that, in general, the expression from the IR-lines method  gives 
higher ICF   values,  and consequently  higher Ne/H values, by about 0.08-0.13\,dex than those obtained using ICFs from 
 \citet{perez07}, \citet{izotov06b}, and from our photoionization models . For
low-excitation objects, we find 
that the high metallicity models by \citet{izotov06b} are in agreement with our
theoretical relation. Finally, the classical ICF approximation yields Ne/H values
in consonance with the ICFs via the IR-lines method.

\subsection{Ne/O vs.\ O/H}

 A very important issue is the study of the relation between the Ne/O
ratio and 
the total oxygen abundance, for which there is not a consensus. \citet{wang08},
using oxygen and neon abundances of a sample of planetary nebulae and
\ion{H}{ii}Rs, found that the Ne/O ratio increases with O/H in both
types of nebulae. A similar result was found by \citet{guseva11} and
\citet{izotov06b} for a large sample of low-metallicity ELGs. 
\citet{perez07}, using a photoionization model grid to derive the Ne ICF
found, for a large sample of SFRs, that the Ne/O behaviour agrees
with the assumption of a constant value for the low metallicity regime
(12+log(O/H)$<$8.2) but shows a slightly decrease for the high metallicity
regime (12+log(O/H)$>$8.2). 
  Moreover, \citet{willner02}, using mid-IR derivations of the neon abundances of 
a sample of \ion{H}{ii}Rs located in M\,33, obtained a flat neon abundance
gradient as a function of the galactocentric radius, as well as a decrement of
the Ne/O ratio when the O/H increases.

\begin{figure}
\centering
\includegraphics[angle=-90,width=9cm]{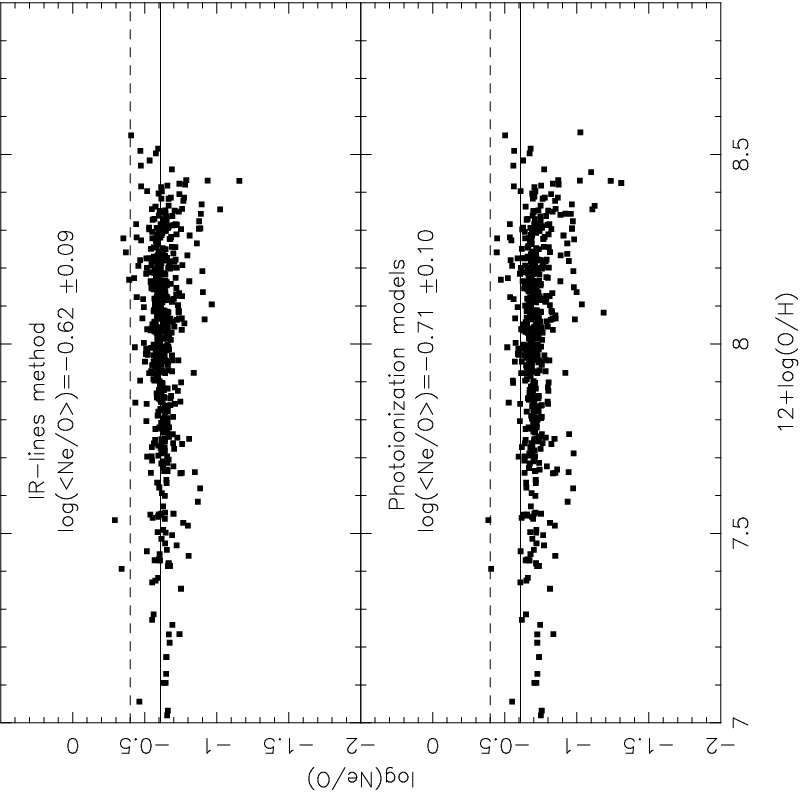}
\caption{Relation between log(Ne/O) and O/H using    the Visible-lines method and 
our two different ICF($\rm  Ne^{++}$) as indicated.  
   Points represent estimations for the objects belonging to the groups A and B. The    solid  line represents the Ne/O
  abundance ratio derived using the oxygen abundance from \citet{allende01}
  and the neon from \citet{grevesse98}, while the    dashed line represents
  the value derived by \citet{drake05} using $Chandra$ X-ray spectra of a
  sample of nearby solar-like stars. The average and the standard deviation of
  log(Ne/O) taking into account low and high  metallicity regimes are
  shown in each plot.} 
\label{f7}
\end{figure}

\begin{figure}
\centering
\includegraphics[angle=-90,width=9cm]{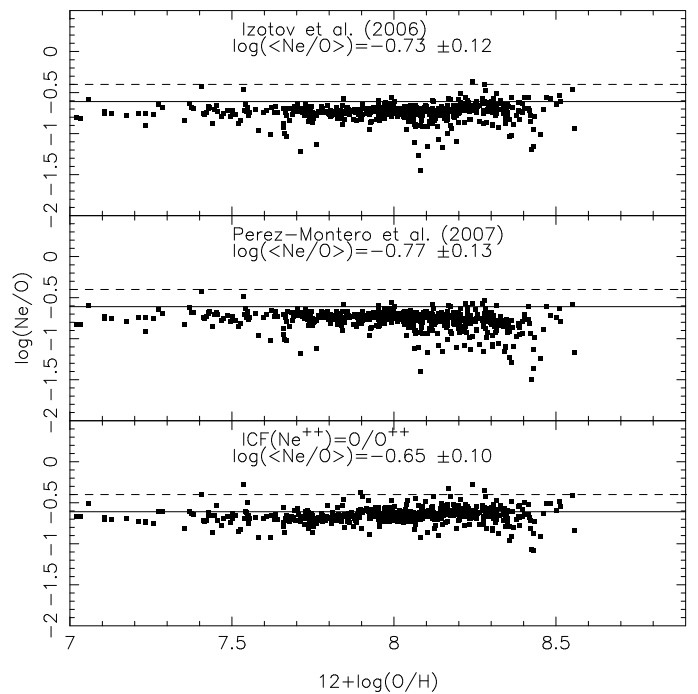}
\caption{As Fig.~\ref{f7} for the ICF($\rm Ne^{++}$) from the literature as
  indicated.} 
\label{f8}
\end{figure}

To investigate this issue, we applied the different ICFs for Ne shown in Fig.~\ref{f5} to
the ionic $\rm Ne^{++}$ abundances estimated using the 
  Visible-lines method to obtain total 
Ne abundances for the compiled sample. We then combined these
values with the total oxygen abundances to estimate Ne/O ratio. 
Figs.\ \ref{f7} and \ref{f8} show these values as a function of 12+log(O/H)
for our ICFs and those from the literature, 
respectively. In these figures we also show the logarithm of the average Ne/O
value derived using the different ICFs, and their standard deviations. 
  These average values are consistent in all the cases with
the adopted solar one. 
From a visual inspection of the distributions of our observational sample in
these diagrams, can be seen that there is a flat relation between the
logarithms of the Ne/O and the O/H ratios for the whole studied metallicity
range, from about 7 to 8.5\,dex. We also perform a
fitting to these data assuming a linear regression 
with slope $a$ and the regression constant $b$, and without taking into account
the individual errors. In Table~\ref{tab2} we list the coefficients of the
fittings. 
We obtain null, positive and negative slopes for the different ICFs. However,
in all the cases these slopes are close to zero. In the worst case,
$a$\,=\,$-$0.08, a metallicity variation equal to the complete metallicity range
(1.5 dex, equivalent to a factor of about 32) implies a variation of the 
Ne/O ratio equal to 0.12 dex (equivalent to a factor of about 1.3). This Ne/O
variation is very similar to the standard deviation of the data. We also
analysed the statistical dependence of Ne/O with O/H using the Spearman's
rank correlation on the data, finding that the Ne/O is statistically constant
independently of the metallicity value. 
Another test was done considering different metallicity regimes. We computed
the average Ne/O value for the high (12+log(O/H)$>$8.2) and the low
(12+log(O/H)$<$8.2) metallicity regime for each considered ICF. In
Table~\ref{tab2} 
these results are listed. None variation of the Ne/O with O/H is noted for
these different regimes and for each ICF. Therefore, we can conclude that
the Ne/O and the metallicity have a flat relation for this sample.
  This flat Ne/O vs. O/H behaviour for the whole range of O/H analysed is a robust
test of nucleosynthesis prediction and supports a very limited (if any) oxygen depletion
in dust.   The oxygen  abundances in grains are poorly known. For example,  \citet{izotov06b}
interpreted that the slight increase of Ne/O with the  metallicity  is due to 
depletion of about 20\% of  O onto grains. Also, \citet{peimbert10} found 
an increasing depletion of O atoms of about 0.1 dex with increasing O/H 
 for   three objects in a large range of  metallicity. 
 From our data we derived a  dispersion of  Ne/O for a fixed O/H value
of  $\sim$0.10 dex.  Therefore, none conclusion can be obtained about the oxygen depletion on dust.

\begin{table*}
\centering
\caption{Coefficients of the linear regressions log(Ne/O)=$a$\,$\times$\,[12+log(O/H)]+$b$ fitted to data plotted in
 Figs.\ \ref{f7} and \ref{f8}. Columns 2 and 3 list the slope and the
 constant coefficient, respectively, together with their formal errors. Columns 4 and 5 are the mean values of
 the log(Ne/O) for the high and low metallicity regimes considered.}
\label{tab2}
\vspace{0.2cm}
\begin{tabular}{@{}llllll@{}}
\hline
 &&& \multicolumn{2}{c@{}}{log($<$Ne/O$>$)}  \\
 \cline{4-5}
\noalign{\smallskip}
ICF($\rm Ne^{++}$)                & \multicolumn{1}{c}{$a$}  &  \multicolumn{1}{c}{$b$}&    \multicolumn{1}{c}{high $Z$} &     \multicolumn{1}{c@{}}{low $Z$}               \\
\hline                                                                             
IR-lines method                         &     +0.02 $\pm 0.01$  &  $-$0.77 $\pm 0.11$   &    $-0.63 \pm0.11$  &  $-0.62 \pm0.08$  \\
Photoionization models                 &  $-$0.02 $\pm 0.01$  &  $-$0.53 $\pm 0.10$   &    $-0.74 \pm0.13$  &  $-0.70 \pm0.09$  \\
\citet{izotov06b}                &     +0.06 $\pm 0.01$ &   $-$1.17 $\pm 0.12$ &    $-0.70 \pm0.12$  &  $-0.75 \pm0.10$  \\ 
P\'erez-Montero                  &  $-$0.08 $\pm 0.01$  &  $-$0.08 $\pm 0.14$  &    $-0.83 \pm0.13$  &  $-0.75 \pm0.13$  \\
ICF($\rm Ne^{++}$)=$\rm O/O^{++}$   &     +0.04 $\pm 0.01$ &   $-$1.06 $\pm 0.12$  &   $-0.64 \pm0.11$  &  $-0.65 \pm0.09$   \\
\hline
 \end{tabular}
\end{table*}

On the other hand, it must be noted that the scatter of the Ne/O increases
with the metallicity, showing the distribution of the data a triangular-like
shape. This result was also found by \citet{kennicutt03} for  \ion{H}{ii}Rs
in M101, NGC\,2403 and dwarf irregular galaxies. They argued that this effect
can reflect high sensitivity to the local radiation field as Ne$^{++}$ and
O$^{++}$ become minor constituents of the nebular material when the O$^{+}$/O
increases, which is equivalent to a metallicity increment. As the metallicity
of the gas material increases the electron temperature decreases, thus, the
Ne$^{++}$ and O$^{++}$ emission-line intensities becomes weak. Therefore, the
increment of scatter going to the higher abundances should be a normal  
behaviour admitting that errors in the measurements of the sensitive lines are
found to be larger (statistically) for higher abundance \ion{H}{ii}Rs. 
In fact the triangular shape mentioned is probably just a product of this.
  We have also tested the dependence of Ne/O on the temperature
fluctuations, since  neon and oxygen have slightly different temperature
coefficients. Using the  earlier result models and computing Ne/H and O/H
following the same procedure  described  in Sect.~\ref{ion}, we found that
log(Ne/O) is $-$0.53 and $-$0.36 for P=0.0 and  0.01, respectively. Thus,
temperature fluctuations would not be  a source of systematic errors in the
Ne/O derivation.

\begin{figure}
\centering
\includegraphics[angle=-90,width=9cm]{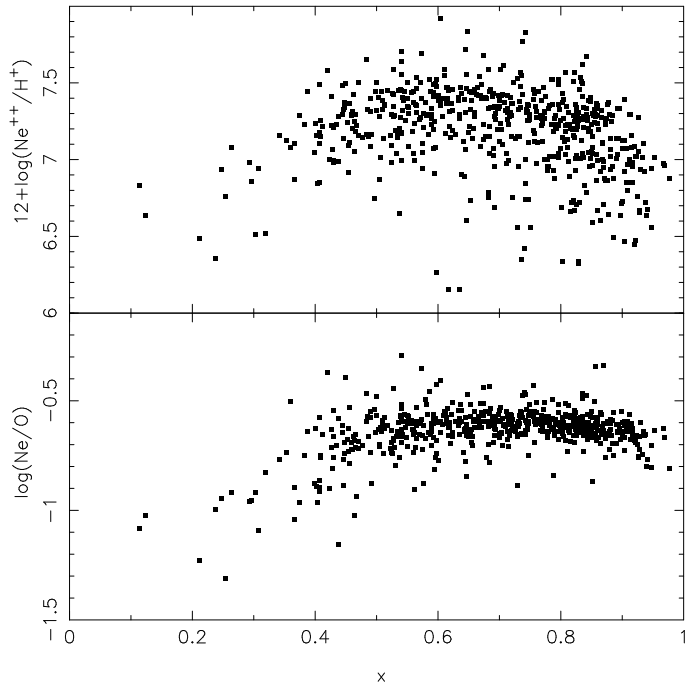}
\caption{  Bottom panel: Relation between log(Ne/O) and  the ionization
 degree x=$\rm O^{++}/(O^{+}+O^{++})$ obtained using  the Visible-lines
 method, the  compiled sample and the ICF for Ne from the IR-lines
 method. Points represent estimations for the objects belonging to the groups
 A and B. Top panel: Relation between estimations of 12+log($\rm
 Ne^{++})/H^{+}$) and x.} 
\label{f11}
\end{figure}

Finally, we  tested the  Ne/O dependence on the ionization degree x=$\rm
O^{++}/(O^{+}+O^{++})$.  To do  that, we  used the  compiled sample to
obtain the    Ne/O abundances via  Visible-lines method and the x values. The
ICF for Ne  from IR-lines method (Eq.~\ref{eq5}) was considered. In
Fig.~\ref{f11} (bottom panel) we can see  that higher Ne/O values are found
for objects with high excitation. This result is independent of the ICF
considered.  Although the neon ICF and O/H ratio decrease with x, the
behaviour of the Ne/O ratio is due to the increment in the $\rm Ne^{++}/H^{+}$
with x, as can be seen in Fig.~\ref{f11} (top panel).

\section{Conclusions}
\label{conc}
In this work we show that the ionic fractions $\rm Ne^{++}/H^{+}$ obtained via
optical direct detections of the electron temperature are underestimated by
  about a factor of 4   in relation to those estimated via mid-infrared emission-lines.
We discussed several possible causes responsible of this discrepancy, including:

(i) Different apertures between optical and mid-IR observations, as was already
suggested by \cite{vermeij02a}. This effect should not be of great 
importance for the most part of the analysed sample, which behaves as point-like
sources. Besides, this effect has minimum effect in a high-excitation ion
such as Ne$^{++}$.

(ii) Differential extinction effects. Mid-IR emission-lines are possibly
tracing a deeper ionization structure than optical ones. 
Besides, due to no reliable measurements of the Hu$\alpha$ hydrogen
recombination line at 10.52\,$\mu$m since it is blended with an H$_2$
line, ionic abundances were derived in relation to H$\beta$ or Br$\beta$
emission-lines. 
We did not find any correlation between the degree of discrepancy and the
inner extinction found from the Balmer decrement.

(iii) Uncertainties in the atomic data related with the infrared emission of both
[\ion{Ne}{ii}] and [\ion{Ne}{iii}]. These emission-lines present some problems
as it is evidenced by 
inspecting the right lower panel of Fig.\ 1 in which observational data and 
photoionization models are shown. According to it the models predict
systematically lower values of the mid-IR [\ion{Ne}{iii}]/[\ion{Ne}{ii}]
emission-line ratio than observations.
This discrepancy was already found by
\cite{morisset04} and \cite{perez09a} and apparently it is not related
with the shape of the used spectral energy distribution in the models.
  The values for the [\ion{Ne}{iii}]$\lambda15.56\mu{\rm m}$
emission-line collision strength found in the literature along decades differ
by about 50\%.
Even when we assume that collision strength for the [\ion{Ne}{iii}]$\lambda15.56\mu{\rm m}$ 
  varies by a factor of two,  it is not enough to conciliate the neon ionic abundance discrepancy.

(iv) Variations of chemical abundances across the nebula. Using a grid of
photoionization models we showed that fluctuations with an amplitude of
0.8\,dex in the total abundance (for both O and Ne) can produce the degree of 
discrepancy between the ionic Ne$^{++}$ abundances derived from optical
and mid-IR emission-lines. 

\medskip

 Using ionic Ne abundances estimations from mid-IR emission-lines, we obtained
an expression for the ICF of the $\rm Ne^{++}$  as a function of the
O$^{++}$/(O$^{+}$+O$^{++}$) ratio. These ICFs obtained through the
use of empirical and theoretical derivations of Ne$^{++}$ from mid-IR lines  
come from a quotient of emissivities and therefore minimize the impact of
points (ii) and (iii) mentioned above, which are more critical for the
abundance in absolute value. We then employed a large sample
of observational spectroscopic data of star-forming regions compiled from the
literature to analyse the dependence  of the Ne/O abundance ratio with
O/H. We found that Ne/O is about constant with O/H ($\rm \log Ne/O \approx -0.70$) for the whole metallicity
range considered.   This result is independent of the approximation 
adopted for the ICF, and in all the cases the average Ne/O estimated for our
sample is consistent with the adopted solar value.

\section*{Acknowledgments}
We are grateful to  the referee   Dr. Steven Willner  for his  useful comments and
suggestions that helped us to substantially clarify and improve the
manuscript. OLD and ACK are grateful to the FAPESP for support under
grant 2009/14787-7 and 2010/01490-3. This work has been partially supported by project AYA2010-21887-C04
of the Spanish National Plan for Astronomy and Astrophysics, and by the project TIC114 
{\em Galaxias y Cosmolog\'\i a} of the
 Junta de Andaluc\'\i a (Spain). R.\~R. thanks to FAPERGs (ARD 11/1758-5) and CNPq).

\label{lastpage}

\end{document}